\documentclass[twocolumn,journal]{IEEEtran}
\IEEEoverridecommandlockouts
\makeatletter
\def\endthebibliography{%
	\def\@noitemerr{\@latex@warning{Empty `thebibliography' environment}}%
	\endlist
}

\usepackage[left=1.33cm,right=1.33cm,top=1.9cm,bottom=4.3cm]{geometry}
\makeatother
\usepackage{cite}
\usepackage{amsthm,amssymb,amsfonts}
\usepackage{amsmath}
\usepackage[linesnumbered,ruled]{algorithm2e}
\SetAlFnt{\small}
\usepackage{graphicx}
\usepackage{xcolor}
\usepackage{array}
\usepackage{verbatim}
\usepackage{multirow}
\usepackage{booktabs}
\usepackage{makecell}
\usepackage{balance}
\usepackage{floatrow}
\usepackage{subfig}
\let\labelindent\relax
\usepackage{enumerate}
\usepackage{enumitem}
\usepackage[font=small]{caption}
\usepackage{comment}
\newcolumntype{P}[1]{>{\centering\arraybackslash}p{#1}}
\usepackage[citecolor=blue]{hyperref}
\hypersetup{draft}

\newtheorem{theorem}{Theorem}

\newtheorem{lemma}{Lemma}
\newtheorem{proposition}{Proposition}
\def\BibTeX{{\rm B\kern-.05em{\sc i\kern-.025em b}\kern-.08em 
		T\kern-.1667em\lower.7ex\hbox{E}\kern-.125emX}}

\usepackage[capitalize]{cleveref}
\usepackage{tikz}
\usepackage{setspace}
\usepackage{changepage}

\begin{document} 
	\title{Unified Network Modeling for Six Cross-Layer Scenarios in Space-Air-Ground Integrated Networks}
	\author{
		\authorblockN{Yalin Liu\authorrefmark{1}}, 
		\authorblockN{Yaru Fu\authorrefmark{1}}, 
		\authorblockN{Qubeijian Wang\authorrefmark{2}}, and \authorblockN{Hong-Ning Dai\authorrefmark{3}}\\
	\authorblockA{
		\authorrefmark{1}Hong Kong Metropolitan University, Email: \{ylliu, yfu\}@hkmu.edu.hk }\\
		\authorrefmark{2}Northwestern Polytechnical University, Email: qubeijian.wang@nwpu.edu.cn\\
		\authorrefmark{3}Hong Kong Baptist University, Email: hndai@ieee.org
	\vspace{-3.5em}}			

\pagestyle{empty}
\maketitle
\thispagestyle{empty}

\begin{abstract}
The space-air-ground integrated network (SAGIN) can enable global range and seamless coverage in the future network. SAGINs consist of three spatial layer network nodes: 1) satellites on the space layer, 2) aerial vehicles on the aerial layer, and 3) ground devices on the ground layer. Data transmissions in SAGINs include six unique cross-spatial-layer scenarios, i.e., three uplink and three downlink transmissions across three spatial layers. For simplicity, we call them \textit{six cross-layer scenarios}. Considering the diverse cross-layer scenarios, it is crucial to conduct a unified network modeling regarding node coverage and distributions in all scenarios. To achieve this goal, we develop a unified modeling approach of coverage regions for all six cross-layer scenarios. Given a receiver in each scenario, its coverage region on a transmitter-distributed surface is modeled as a spherical dome. Utilizing spherical geometry, the analytical models of the spherical-dome coverage regions are derived and unified for six cross-layer scenarios. We conduct extensive numerical results to examine the coverage models under varying carrier frequencies, receiver elevation angles, and transceivers' altitudes. Based on the coverage model, we develop an algorithm to generate node distributions under spherical coverage regions, which can assist in testing SAGINs before practical implementations. 
\end{abstract}
\vspace{-0.5cm}





\section{Introduction}
\vspace{-0.1cm}

The increasing deployment of aerial vehicles and satellites has great potential to support seamless coverage and global-range communications. 3GPP has planned numerous standardization study items to integrate non-terrestrial networks into the upcoming 6G network~\cite{3GPPTR38863}. Several leading satellite providers, such as SpaceX and Lynk, are also establishing partnerships with mobile network operators~\cite{23GSA}. By integrating ground, air, and space networks, the space-air-ground integrated network (SAGIN) is a crucial technology to enable the future network. SAGINs consist of three spatial layer nodes: 1) satellites on the space layer, 2) aerial vehicles on the aerial layer, and 3) ground devices on the ground layer~\cite{3GPPTR38811}. Data transmissions in SAGINs include six unique cross-layer scenarios, i.e., three uplink and three downlink transmissions across space, air, and ground layers, respectively. 

Implementing the six cross-layer scenarios in SAGINs is a complex process that requires meticulous planning and testing~\cite{23GSA,3GPPTR38811}. To facilitate this process, it is crucial to conduct robust theoretical research on modeling the network nodes of SAGINs under these cross-layer scenarios. In each scenario, given a receiver, the transmitter(s) within its coverage (or observation) region can establish connections, thereby forming the network topology. In this context, the receiver's coverage and the distribution of transmitters are two critical factors in modeling the network topology, i.e., nodes. It is worth noting that the distribution of transmitters is constrained by the boundaries of the receiver's coverage region. Therefore, accurately modeling the node coverage regions in SAGINs is fundamental and significant.

Some previous studies have considered the global range coverage and node distributions of SAGINs. The authors in~\cite{24iotleo,23BPPsagin} modeled satellites distributed over visible spherical areas with distribution boundaries observable by ground nodes. In~\cite{21twcsg,zhang2021stochastic}, the authors examined the deployment of aerial vehicles following the Poisson cluster process to assist space-to-air/ground transmissions. Although~\cite{23BPPsagin,24iotleo,21twcsg,zhang2021stochastic} modeled the distributions of satellites and aerial vehicles,
they did not provide specific coverage models for network nodes in SAGINs and only considered downlink transmission scenarios, i.e., space-to-ground or space-to-air. Some related studies have considered uplink transmissions in SAGINs, such as ground-to-space~\cite{al2022optimal} and ground-to-air~\cite{liu2023TVT} links. These studies modeled the coverage regions of satellites or aerial vehicles as two-dimensional circles or hexagonal cells, which are oversimplified and not applicable to SAGINs. In~\cite{talgat2024stochastic}, a more practical coverage model for ground-to-space scenarios was proposed, where ground users are distributed on a spherical Earth surface covered by satellites. However, \cite{talgat2024stochastic} used a fixed beamwidth for satellites, whereas practical beamwidths vary with different antenna configurations. The authors in \cite{liu2024} developed a spherical coverage model that considers changing beamwidths with frequencies, but it only applies to uplink SAGIN transmissions. In summary, although~\cite{al2022optimal,liu2023TVT,talgat2024stochastic,liu2024} models nodes and coverage in uplink transmission scenarios, their approaches cannot be applied to diverse cross-layer scenarios in SAGINs.

\begin{figure}[t]
	\centering
	\captionsetup[subfigure]{oneside,margin={0cm,0cm}}
	\subfloat[Uplink scenario $\mathbb{S}_{i,\mathrm{u}}$]{\includegraphics[width=3.1cm]{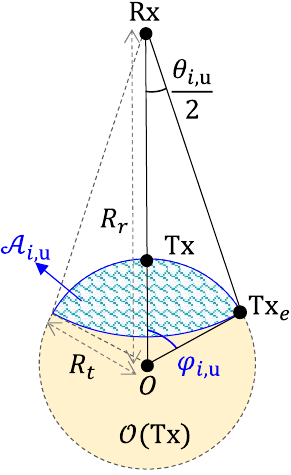}\label{subfig: uplink}}	\hfil 
	\captionsetup[subfigure]{oneside,margin={0cm,0cm}}
	\subfloat[Downlink scenario $\mathbb{S}_{i,\mathrm{d}}$]{\includegraphics[width=3.15cm]{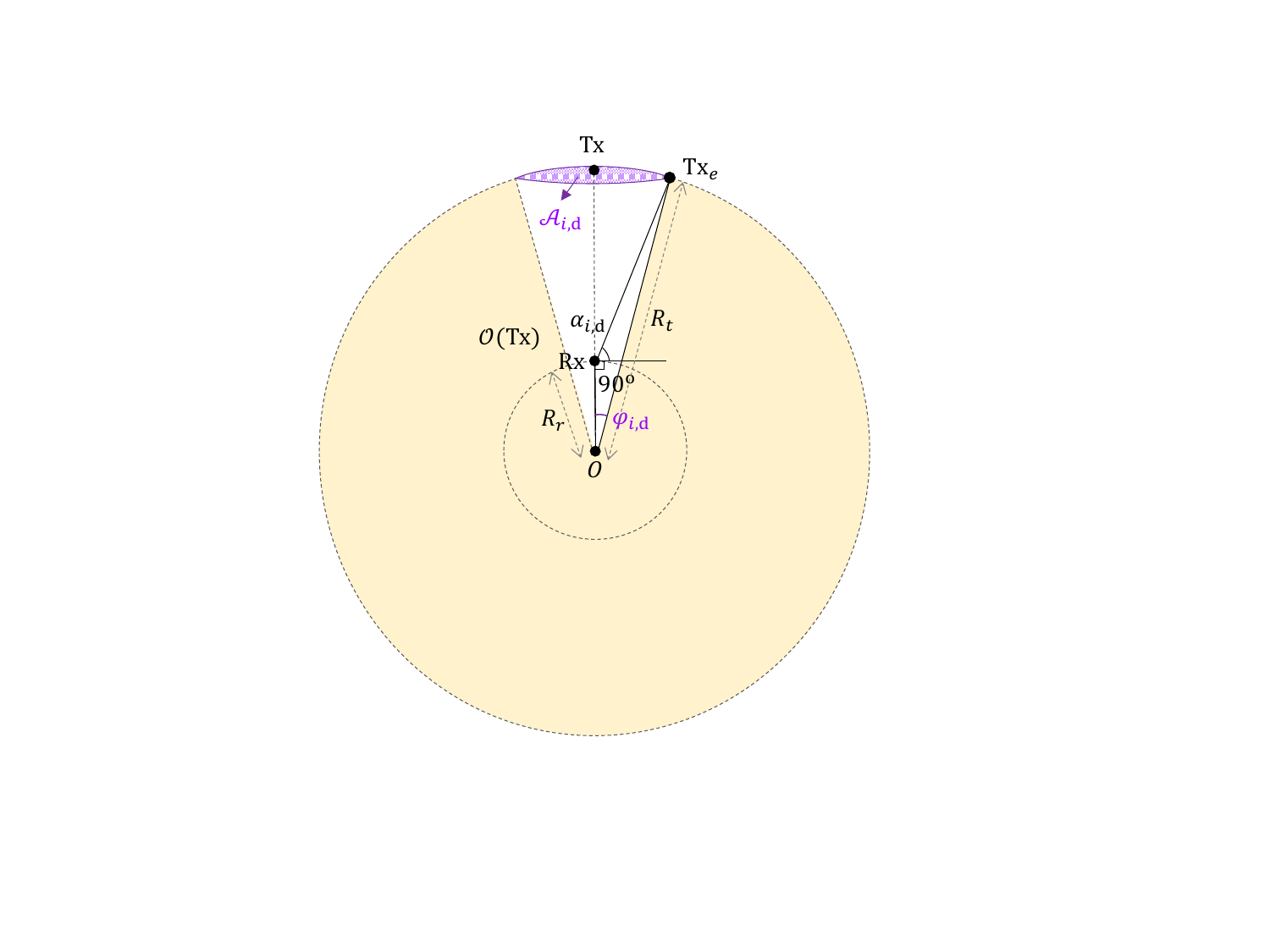}\label{subfig: downlink}}	\hfil 
	\caption{Spherical geometry in six transmission scenarios ($\mathbb{S}_{i,j}(\forall i \in \{1,2,3\}, j\in\{\mathrm{u},\mathrm{d}\})$) of SAGINs, where $\mathrm{Rx},\mathrm{Tx},\mathrm{Tx}_e,\mathcal{O}(\mathrm{Tx})$ are the receiver, the transmitter, the edge point on $\mathcal{A}_{i,j}$, and the sphere surface on which all transmitters with the same altitude distributed.}
	\label{fig: system}
	\vspace{-0.2cm}
\end{figure}

Considering diverse cross-layer scenarios, network modeling in SAGINs is a complex task. First, it is a significant challenge to model coverage regions and node distributions in SAGINs due to the changing spatial relationships in cross-layer scenarios. Second, accurately capturing the diverse configurations across different cross-layer scenarios in SAGINs is difficult. Therefore, it demands a new theoretical framework to consider various antenna settings and transmission parameters, including the altitudes and antennae of satellites and aerial vehicles. To address these challenges, \textit{we present a unified analytical model for coverage regions with consideration of diverse transmission settings. We also provide comprehensive numerical results based on this model. Based on the coverage model, we develop an algorithm to generate the transmitter distribution 
	for all scenarios}. Our contributions are summarized below:

\begin{itemize}
	\item We present a comprehensive system model of SAGINs in six cross-layer scenarios. A spherical coordinate system is used to represent the spatial coordinates of cross-layer transceivers. Diverse 
	antenna configurations and parameters are considered to cater to all scenarios. 
	\item We develop a unified analytical model of spherical coverage regions in six cross-layer scenarios. The analytical formulas of coverage regions are derived and unified for all six scenarios. Numerical results are conducted to examine the coverage models under varying carrier frequencies, receiver's elevation angles, and transceivers' altitudes. 
	\item We design an algorithm to generate node distributions following stochastic point processes under spherical coverage regions. Simulation results of node distributions are plotted across six cross-layer scenarios. These results offer valuable insights that can assist in the network modeling and testing of SAGINs prior to actual deployment.
\end{itemize}


In the next, 
\cref{sec: system} models six cross-layer scenarios.~\cref{sec: modelanalysis} presents the unified coverage model. 
\cref{sec: distri} gives the distribution algorithm.~\cref{sec: con} concludes this paper.
\vspace{-0.5cm}

\section{System Model}
\label{sec: system}
\vspace{-0.1cm}

\subsection{Six Cross-Layer Scenarios} 
\vspace{-0.1cm}
We consider six cross-layer scenarios in SAGINs as follows. 
\begin{itemize}
	\item $\mathbb{S}_{1,\mathrm{u}}$: the ground-to-air (G2A) uplink scenario. 
	\item $\mathbb{S}_{2,\mathrm{u}}$: the air-to-space (A2S) uplink scenario. 
	\item $\mathbb{S}_{3,\mathrm{u}}$: the ground-to-space (G2S) uplink scenario. 
	\item $\mathbb{S}_{1,\mathrm{d}}$: the air-to-ground (A2G) downlink scenario. 
	\item $\mathbb{S}_{2,\mathrm{d}}$: the space-to-air (S2A) downlink scenario. 
	\item $\mathbb{S}_{3,\mathrm{d}}$: the space-to-ground (S2G) downlink scenario. 
\end{itemize}
For convenience, we use the subscripts $i \in \{1,2,3\}, j\in\{\mathrm{u},\mathrm{d}\}$, to indicate the above six scenarios. 
As shown in~\cref{fig: system}, in three uplink scenarios, 
the receiver (i.e., aerial vehicles/satellites) uses a single-beam directional antenna to cover transmitters~\cite{3GPPTR38811}. Let $\theta_{i,\mathrm{u}}$ be the 3-dB beamwidth of the receiver in $\mathbb{S}_{i,\mathrm{u}}$. For a normalized antenna pattern~\cite[Eq. (5.3b)]{maral2020satellite}, $\theta_{i,\mathrm{u}}$ is given by
\vspace{-0.4cm}

{\small
\begin{align}
	&\forall i \in \{1,2,3\}: 
	\theta_{i,\mathrm{u}}=\frac{\kappa_{i,\mathrm{u}}c}{f_{i,\mathrm{u}}\mathrm{D}_{i,\mathrm{u}}}(\mathrm{degrees}),
	\label{eq: theataG}
\end{align}%
}%
\vspace{-0.5cm}

\noindent where $c$ is the light speed, $f_{i,\mathrm{u}}$ is the carrier frequency, $\kappa_{i,\mathrm{u}}$ is the antenna illumination coefficient, and $\mathrm{D}_{i,\mathrm{u}}$ is the diameter of the normalized reflector antenna. 
In three downlink scenarios, 
each receiver (i.e., aerial vehicle/ground device) is configured with the auto-tracking antenna to position its transmitter~\cite{3GPPTR38811}. The receiver 
in $\mathbb{S}_{i,\mathrm{d}}$ has the minimum elevation angle, denoted by $\alpha_{i,\mathrm{d}}$. 
\vspace{-0.4cm}

\begin{figure}[t]
	\centering
	\includegraphics[width=5.3cm]{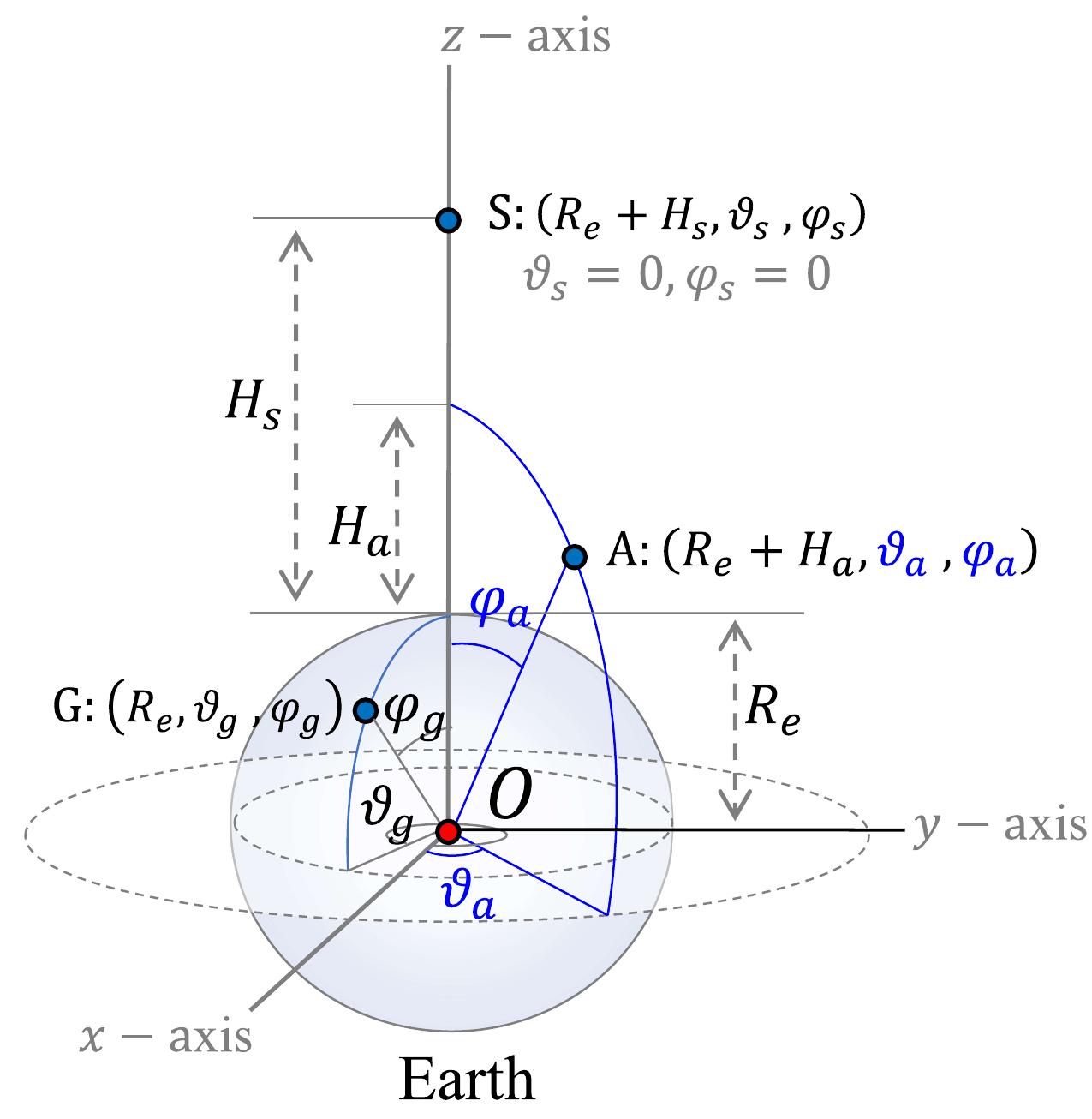}\hfil 
	\caption{Spherical coordinate system in SAGINs, where $\mathrm{G}$, $\mathrm{A}$, and $\mathrm{S}$ represent three reference nodes on ground, air, and space.}
	\label{fig: geo_scs}
	\vspace{-0.2cm}
\end{figure}

\subsection{Spherical Coordinate System} 
\vspace{-0.1cm}
The earth's surface (i.e., the ground) is approximated as \textit{a spherical surface} with the radius $R_e$ and the earth center $O$. In SAGIN, three layers of nodes are distributed around the earth at different altitudes, i.e., $\{0,H_{\mathrm{a}},H_{\mathrm{s}}\}$ for the nodes on ground, air, and space, respectively. To model the spatial locations of all nodes, we build a spherical coordinate system by letting the earth center $O$ be the original point and the orientation from $O$ to any reference direction as the zenith direction, as shown in Fig.~\ref{fig: geo_scs}. In our coordinate system, each node has a 
3D polar coordinate represented by $(r, \vartheta, \varphi)$, where $r$ is the distance 
to $O$, $\vartheta$ and $\varphi$ are the azimuth angle and the polar angle of the node, respectively. 
\vspace{-0.4cm}

\subsection{Parameter Configurations}
\vspace{-0.1cm}
All numerical results in this paper adopts the practical parameter settings in LTE-supported aerial vehicles~\cite{173GPP} (for low altitude platforms), non-terrestrial networks~\cite{3GPPTR38863} (for high altitude platforms and satellites), and satellite communications~\cite{maral2020satellite}. In three uplink scenarios, the antenna configurations at receivers are: the illumination coefficients $\{\kappa_{1,\mathrm{u}},\kappa_{2,\mathrm{u}},\kappa_{3,\mathrm{u}}\}=\{70,70,70\}$, the antenna diameters $\{\mathrm{D}_{1,\mathrm{u}},\mathrm{D}_{2,\mathrm{u}},\mathrm{D}_{3,\mathrm{u}}\}=\{0.2\mathrm{m},4\mathrm{m},4\mathrm{m}\}$. In the three downlink scenarios, the receivers adopt the configurations:
the minimum elevation angles  $\{\alpha_{1,\mathrm{d}},\alpha_{2,\mathrm{d}},\alpha_{3,\mathrm{d}}\}=5\sim 30\mathrm{degrees}$~\cite{3GPPTR381015} (Chapter 4.5). In addition, the scenarios $\mathbb{S}_{1,j} (\forall j\{\mathrm{u},\mathrm{d}\})$ use $\{f_{1,j}\}=300\mathrm{MHz}\sim 2.4\mathrm{GHz}$ and $H_{\mathrm{a}}=1\mathrm{km}\sim 50\mathrm{km}$ for low altitude platforms~\cite{173GPP} and high altitude platforms~\cite{3GPPTR38863}. The other scenarios $\mathbb{S}_{2,j} (\forall j\{\mathrm{u},\mathrm{d}\})$ and $\mathbb{S}_{3,j} (\forall j\{\mathrm{u},\mathrm{d}\})$ use $\{f_{2,j},f_{2,j}\}= 2\mathrm{GHz}\sim 40\mathrm{GHz}$ and $H_{\mathrm{s}}=500\mathrm{km}\sim 35786\mathrm{km}$ for non-terrestrial networks~\cite{3GPPTR38863} and satellite communications~\cite{maral2020satellite}. The satellite altitude $H_{\mathrm{s}}$ crosses three orbits, i.e., $500\mathrm{km}\sim2000\mathrm{km}$ in LEO, $2000\mathrm{km}\sim35786\mathrm{km}$ in MEO, and $35786\mathrm{km}$ in GEO. The earth radius is $R_e=6371\mathrm{km}$.
\vspace{-0.6cm}

\section{Unified Coverage Modeling}
\label{sec: modelanalysis}


\vspace{-0.1cm}
\subsection{Spherical Coverage Region}
\vspace{-0.1cm}
For each cross-layer scenario $\mathbb{S}_{i,j}$ ($ i \in \{1,2,3\},j\in\{\mathrm{u},\mathrm{d}\}$), a typical transmitter-receiver pair is denoted by $\{\mathrm{Tx},\mathrm{Rx}\}$ and their polar coordinates are represented by $\{(R_t,\vartheta_t,\varphi_t),(R_r,\vartheta_r,\varphi_r)\}$, where $R_t=R_e+H_t,R_r=R_e+H_r$ and $H_t,H_r$ are the receiver's altitude and the receiver's altitude, respectively. As shown in Fig.~\ref{fig: system}, all transmitters located at the same altitude $H_\mathrm{t}$ are distributed on a spherical surface (denoted by $\mathcal{O}(\mathrm{Tx})$) centering at $O$ with the radius $R_t$. Let $\mathcal{A}_{i,j}$ ($\forall i \in \{1,2,3\},j\in\{\mathrm{u},\mathrm{d}\}$) be the coverage (or observation) region of the receiver $\mathrm{Rx}$ on transmitters distributed on $\mathcal{O}(\mathrm{Tx})$. 
Fig.~\ref{fig: system}\subref{subfig: uplink} shows the spherical geometry in three uplink scenarios $\mathbb{S}_{i,\mathrm{u}}$ ($\forall i \in \{1,2,3\}$), where the beam of the receiver $\mathrm{Rx}$ is directed to the earth center $O$. Given the transmitters distributed on $\mathcal{O}(\mathrm{Tx})$, $\mathcal{A}_{i,\mathrm{u}}$ is the intersection between the receiver's beam and $\mathcal{O}(\mathrm{Tx})$. Fig.~\ref{fig: system}\subref{subfig: downlink} shows the spherical geometry in three downlink scenarios $\mathbb{S}_{i,\mathrm{d}}$ ($\forall i \in \{1,2,3\}$), the receiver has the minimum elevation angle $\alpha_{i,\mathrm{d}}$ to observe the transmitters. Given the transmitters distributed on $\mathcal{O}(\mathrm{Tx})$, $\mathcal{A}_{i,\mathrm{d}}$ is the intersection between the receiver's observation angle (i.e., $1-\alpha_{i,\mathrm{d}}$) and $\mathcal{O}(\mathrm{Tx})$. Overall, \textit{for all six scenarios, $\mathcal{A}_{i,j}$ can be modeled as a spherical dome centering at $(R_t,\vartheta_r,\varphi_r)$.} Let $\mathrm{Area}(\cdot)$ denote the area of a region. Using the integral of the point $\left(R_t,\vartheta_t,\varphi_t\right)$ on $\mathcal{A}_{i,j}$, we have~\cref{pro1}.
\vspace{-0.4cm}

\begin{lemma}
	For six cross-layer scenarios $\mathbb{S}_{i,j}$ ($\forall i \in \{1,2,3\},j\in\{\mathrm{u},\mathrm{d}\}$), 
	the coverage region area of $\mathrm{Rx}$ on $\mathrm{Tx}$ 
	is given by
	\vspace{-0.4cm}
	
	{\small
		\begin{align*}
			\mathrm{Area}(\mathcal{A}_{i,j})&=
			\int_{0}^{\varphi_{i,j}}\int_{0}^{2\pi} R_t^2\sin\varphi_t\mathbf{d}\vartheta_t\mathbf{d}\varphi_t
			=2\pi R_t^2 (1-\cos(\varphi_{i,j})),
		\end{align*}%
	}%
	\vspace{-0.4cm}
	
	\noindent where $\varphi_{i,j}$ is the vertex angle of $\mathcal{A}_{i,j}$ as shown in~\cref{fig: system}. 
	\label{pro1}
\end{lemma}%
\noindent In practice, 
$\mathcal{A}_{1,j}$ tends to be a horizontal circular region. This is because of the much shorter altitude of aerial vehicles compared with the earth's radius. Even though, our model can accurate characterize $\mathcal{A}_{1,j}$. Meanwhile, our model can also be used to find the area of the horizontal circular region $\mathcal{A}_{1,\mathrm{u}}$, i.e., $\pi R_t^2 \varphi_{1,j}^2$. 
To further evaluate $\mathrm{Area}(\mathcal{A}_{i,j})$ in~\cref{pro1}, we need to find $\varphi_{i,j}$. 
According to Fig.~\ref{fig: system}, we can calculate $\varphi_{i,j}$ by~\cref{cor1}.
\vspace{-0.4cm}

\begin{adjustwidth}{0cm}{0.1cm}
\begin{proposition}
	For six cross-layer scenarios $\mathbb{S}_{i,j}$ ($\forall i \in \{1,2,3\},j\in\{\mathrm{u},\mathrm{d}\}$), 
	$\varphi_{i,j}$ can be calculated by
	\vspace{-0.4cm}
	
	{\small
	\begin{align*}
		&\varphi_{i,\mathrm{u}}=\left\{\begin{matrix}
			\arccos\left (\delta_{i,\mathrm{u}}\right ),\mathrm{ if}\frac{\theta_{i,\mathrm{u}}}{2}\leq \arcsin\left(\frac{R_t}{R_r}\right).\\
			\arccos\left(\frac{R_t}{R_r}\right),\mathrm{ if}\frac{\theta_{i,\mathrm{u}}}{2}> \arcsin\left(\frac{R_t}{R_r}\right).
		\end{matrix}\right.\varphi_{i,\mathrm{d}}=\arccos\left(\delta_{i,\mathrm{d}}\right),
	\end{align*}%
	}%
	\vspace{-1.2cm}
	
	{\small
		\begin{align*}
			&\delta_{i,\mathrm{u}}= \frac{R_r}{R_t}\sin^2\left(\frac{\theta_{i,\mathrm{u}}}{2}\right) 
			+\cos\left (\frac{\theta_{i,\mathrm{u}}}{2} \right )\sqrt{1-\frac{R_r^2}{R_t^2}\sin^2\left(\frac{\theta_{i,\mathrm{u}}}{2}\right)}, \\ 		
			&\delta_{i,\mathrm{d}}=\frac{R_r}{R_t}\cos^2\alpha_{i,\mathrm{d}}+ \sin\alpha_{i,\mathrm{d}} \sqrt{1-\frac{R_r^2}{R_t^2}\cos^2\alpha_{i,\mathrm{d}}},
		\end{align*}%
	}%
	\vspace{-0.4cm}
	
	\noindent 
	where $R_t=\{R_e,R_e+H_{\mathrm{a}},R_e,R_e+H_{\mathrm{a}},R_e+H_{\mathrm{s}},R_e+H_{\mathrm{s}}\}$ and $R_r=\{R_e,R_e+H_{\mathrm{a}},R_e,R_e+H_{\mathrm{a}},R_e+H_{\mathrm{a}},R_e+H_{\mathrm{s}},R_e+H_{\mathrm{s}}\}$ in six scenarios $\mathbb{S}_{1,\mathrm{u}},\mathbb{S}_{2,\mathrm{u}},\mathbb{S}_{3,\mathrm{u}},\mathbb{S}_{1,\mathrm{d}},\mathbb{S}_{2,\mathrm{d}},\mathbb{S}_{3,\mathrm{d}}$, respectively.
	\label{cor1}
\end{proposition}
\end{adjustwidth}
\begin{proof}
	The proof is given in Appendix~A.
\end{proof}
\noindent In~\cref{cor1}, the vertex angles $\varphi_{i,\mathrm{u}}$ 
in three uplink scenarios 
have two different expressions, depending on two comparative angles (i.e., $\theta_{i,\mathrm{u}}/2$ and $\arcsin(R_t/R_r)$) for each scenario. The beamwidth $\theta_{i,\mathrm{u}}$ can be calculated by \eqref{eq: theataG}. The numerical ranges of two comparative angles for three scenarios are analyzed in our previous work~\cite{liu2024}.  
Referring to Fig.~6 in~\cite{liu2024}, $\theta_{1,\mathrm{u}}/2\leq \arcsin\left (R_t/R_r \right )$ and $\varphi_{1,\mathrm{u}}=\arccos\left (\delta_1\right )$ always holds in the general frequency range of $\mathbb{S}_{1,\mathrm{u}}$, i.e., $f_{1,\mathrm{u}}>300$MHz. Meanwhile, $\theta_{2,\mathrm{u}}/2\leq \arcsin\left (R_t/R_r\right ),\theta_{3,\mathrm{u}}/2\leq \arcsin\left (R_t/R_r \right )$ always hold in the general settings of non-terrestrial networks. 
As a result, the expression of $\varphi_{i,\mathrm{u}}$ in three uplink scenarios can be reduced to $\varphi_{i,\mathrm{u}}=\arccos\left (\delta_{i,\mathrm{u}}\right )$ under general value ranges of $f_{i,\mathrm{u}},H_r,H_t$. Substituting 
\cref{cor1} into~\cref{pro1}, we have~\cref{the1}. 
\vspace{-0.4cm}

\begin{adjustwidth}{0cm}{0.1cm}
\begin{theorem}
	For six cross-layer scenarios $\mathbb{S}_{i,j}$ ($\forall i \in \{1,2,3\},j\in\{\mathrm{u},\mathrm{d}\}$), under general value ranges of $f_{i,\mathrm{u}},\alpha_{i,\mathrm{d}},H_r$ and $H_t$, the coverage region area of $\mathrm{Rx}$ on $\mathrm{Tx}$ 
	can be calculated by
	\vspace{-0.4cm}
	
	{\small
		\begin{align*}
			&\mathrm{Area}(\mathcal{A}_{i,j})=2\pi R_t^2 (1-\cos(\varphi_{i,j}))=
			2\pi R_t^2 (1-\delta_{i,j}),
		\end{align*}%
	}%
	\vspace{-0.6cm}
	
	\noindent where $\delta_{i,\mathrm{u}},\delta_{i,\mathrm{d}}$ are given in~\cref{pro1}.
	\label{the1}
\end{theorem} 
\end{adjustwidth}
\vspace{-0.6cm}

\subsection{Numerical Analysis}
\label{subsec: numerical}
\vspace{-0.1cm}

\begin{figure*}[htbp]
	\centering
	\subfloat[$\mathbb{S}_{1,\mathrm{u}}$(G2A)]{\includegraphics[width=4.35cm]{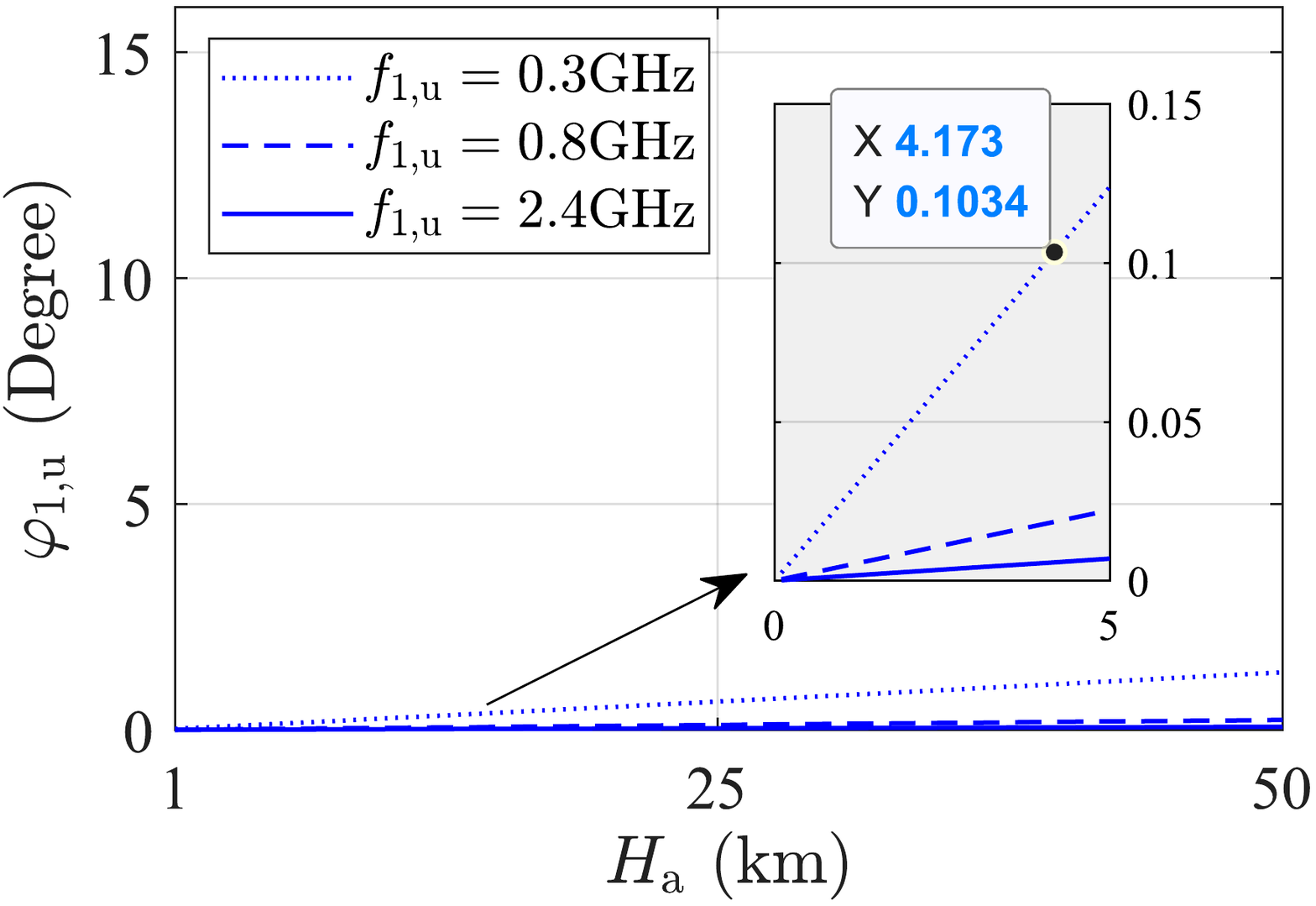}\label{fig: up1_phi_H}}	\hfil 
	\subfloat[$\mathbb{S}_{1,\mathrm{u}}$(G2A)]{\includegraphics[width=4.25cm]{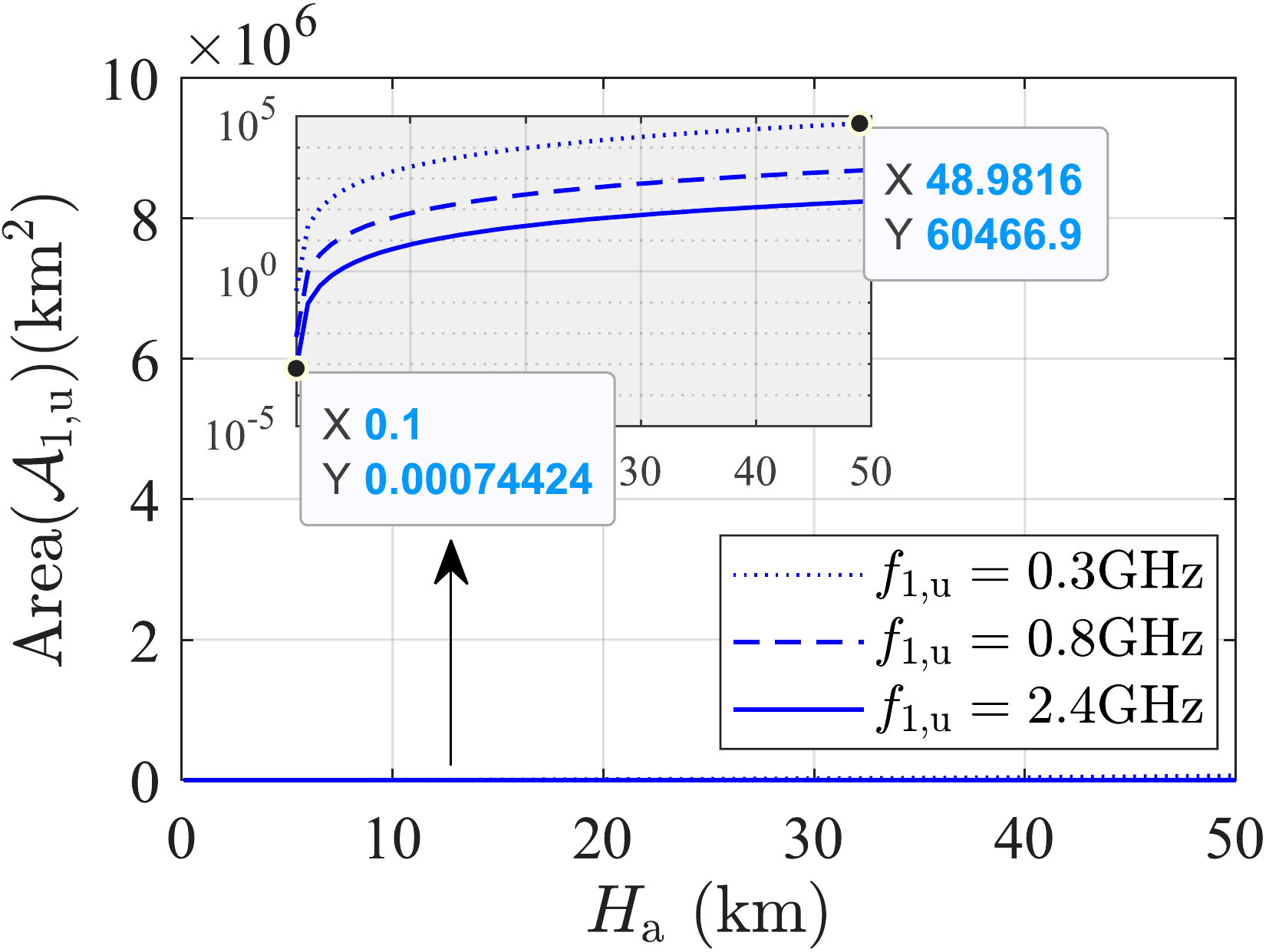}\label{fig: up1_area_H}}	\hfil   
	\subfloat[$\mathbb{S}_{1,\mathrm{d}}$(A2G)]{\includegraphics[width=4.05cm]{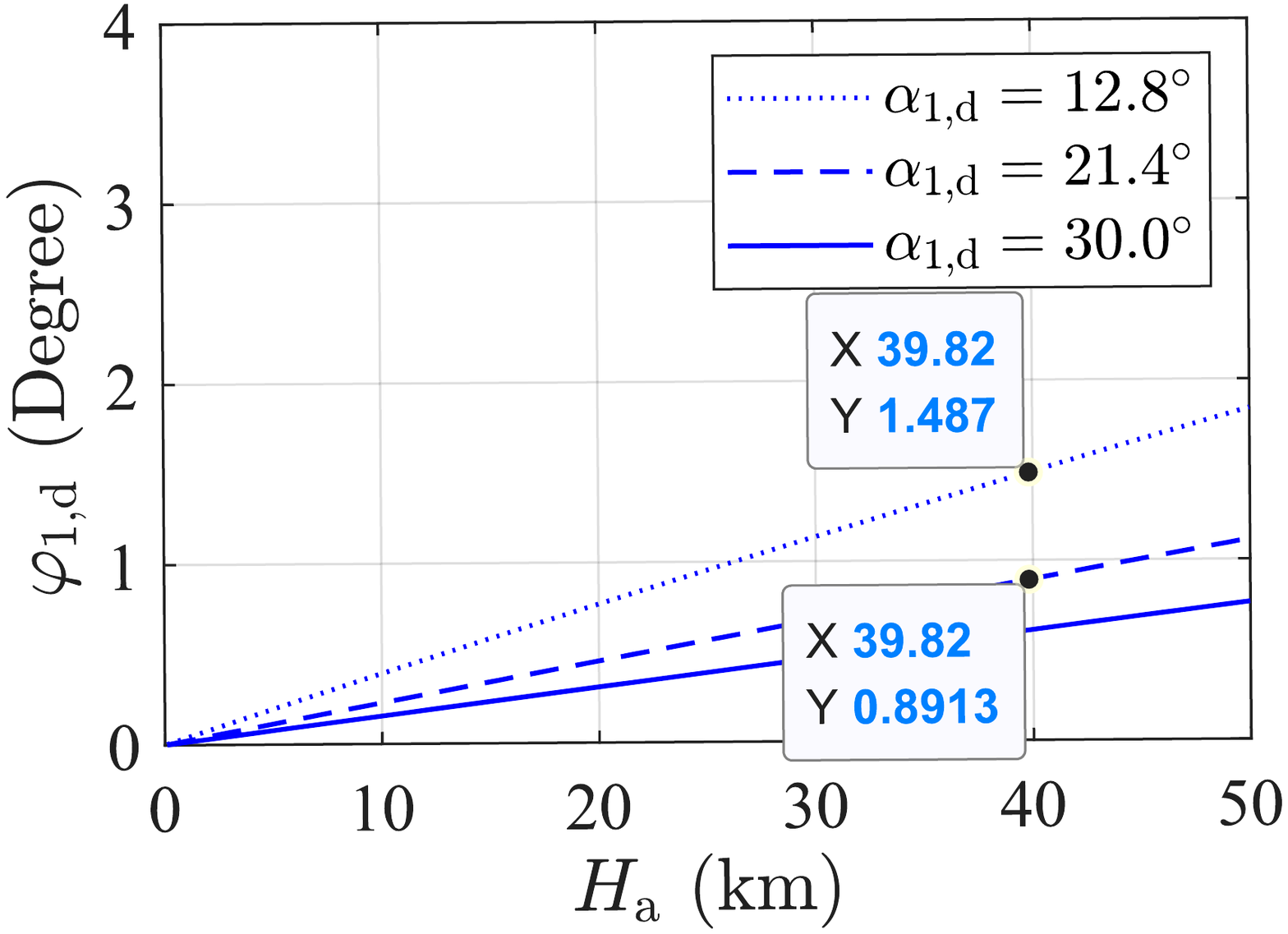}\label{fig: down1_phi_H}}	\hfil 
	\subfloat[$\mathbb{S}_{1,\mathrm{d}}$(A2G)]{\includegraphics[width=4.45cm]{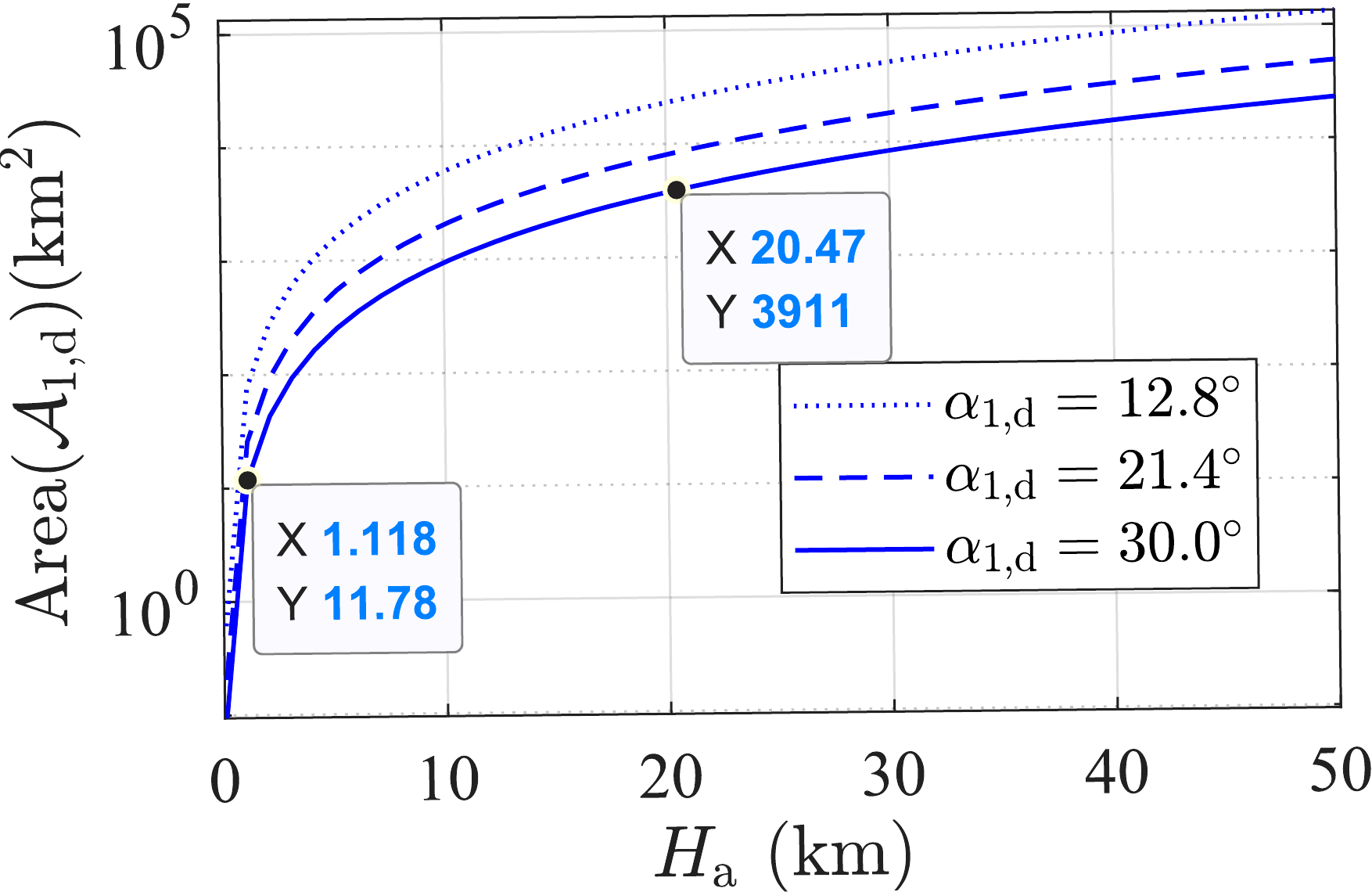}\label{fig: down1_area_H}}	\hfil  
	\\[-2ex]
	\subfloat[$\mathbb{S}_{2,\mathrm{u}}$(A2S)]{\includegraphics[width=4.5cm]{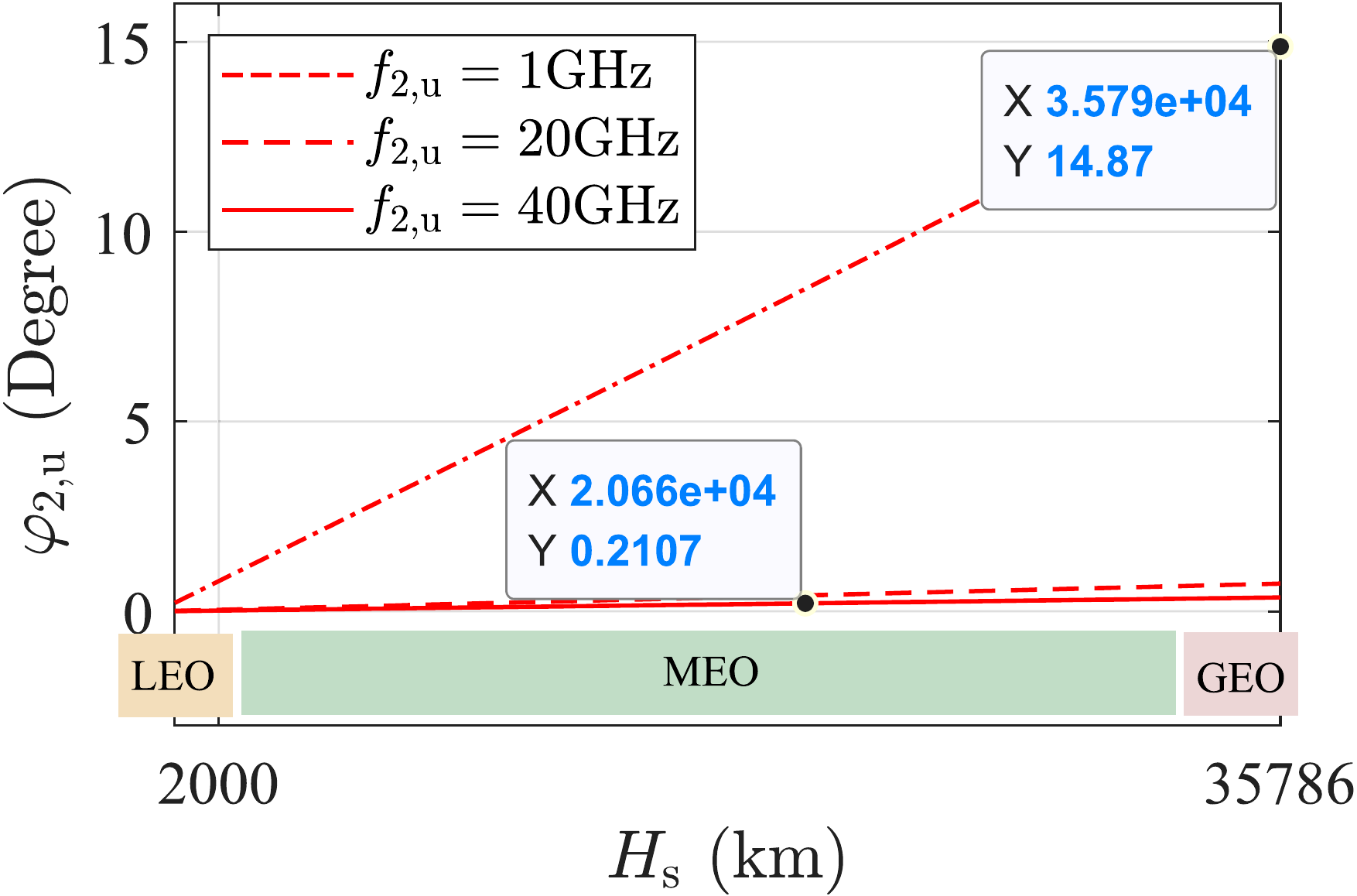}\label{fig: up2_phi_H}}	\hfil
	\subfloat[$\mathbb{S}_{2,\mathrm{u}}$(A2S)]{\includegraphics[width=4.3cm]{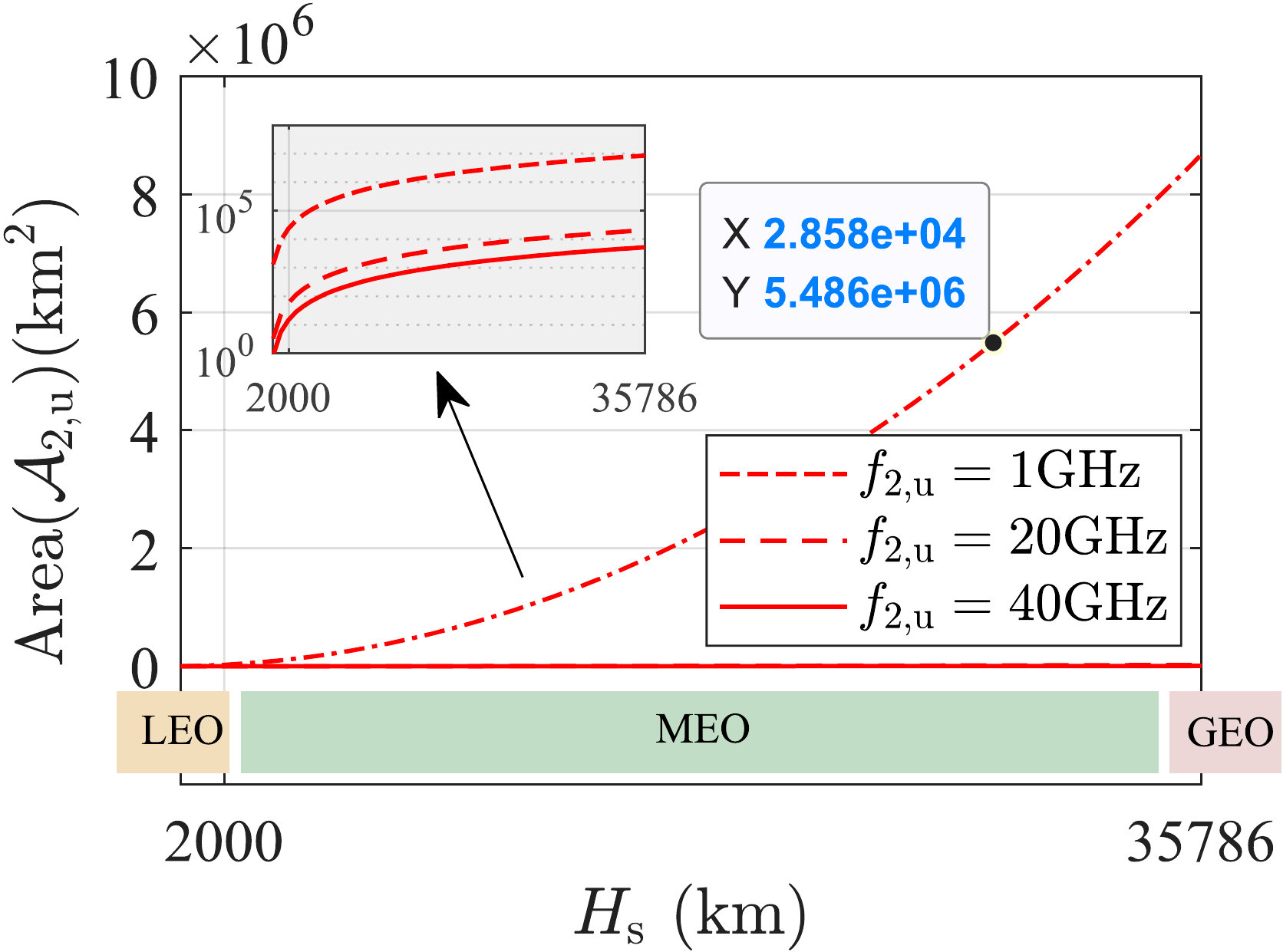}\label{fig: up2_area_H}}	\hfil	
	\subfloat[$\mathbb{S}_{2,\mathrm{d}}$(S2A)]{\includegraphics[width=4.35cm]{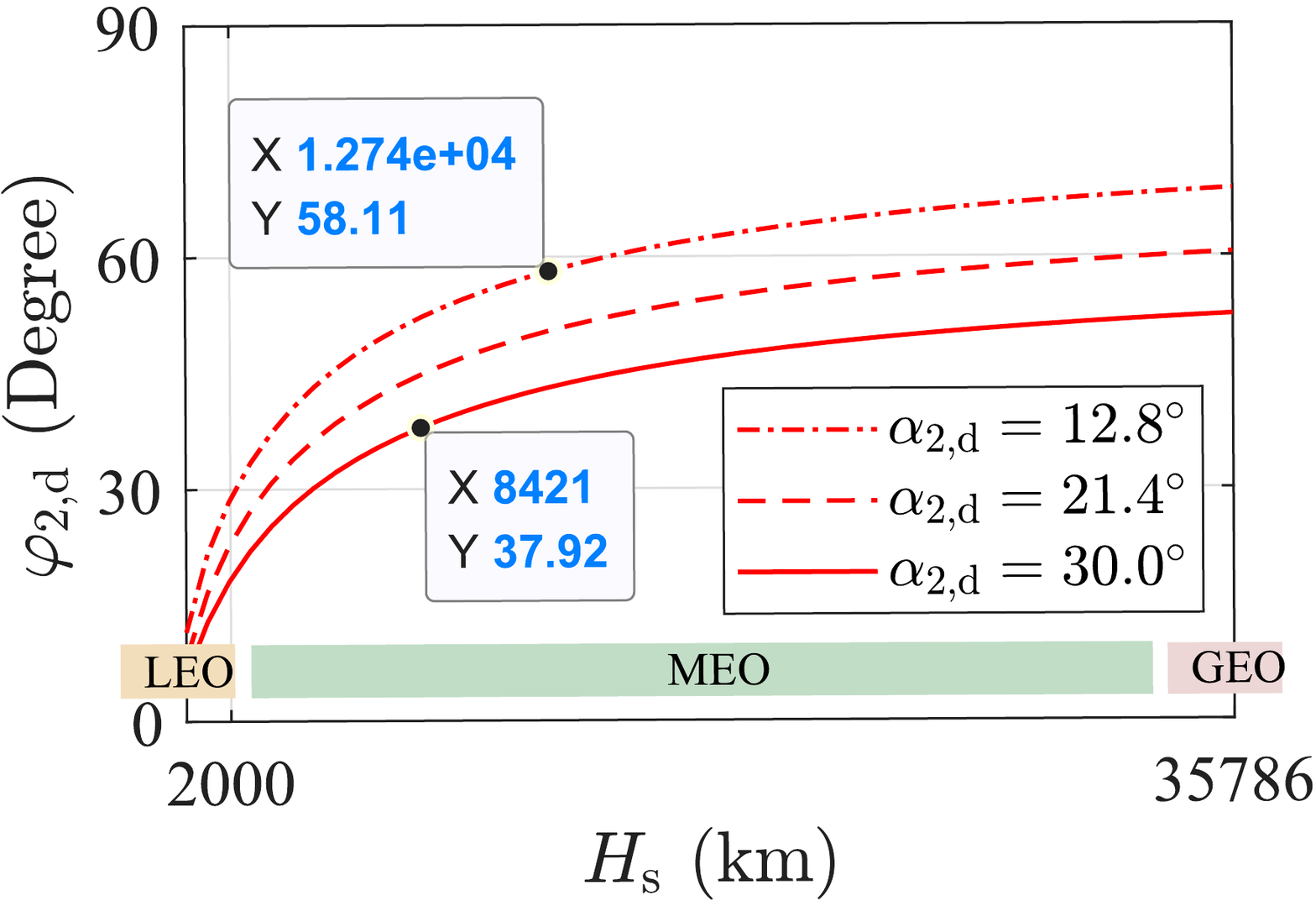}\label{fig: down2_phi_H}}	\hfil
	\subfloat[$\mathbb{S}_{2,\mathrm{d}}$(S2A)]{\includegraphics[width=4.65cm]{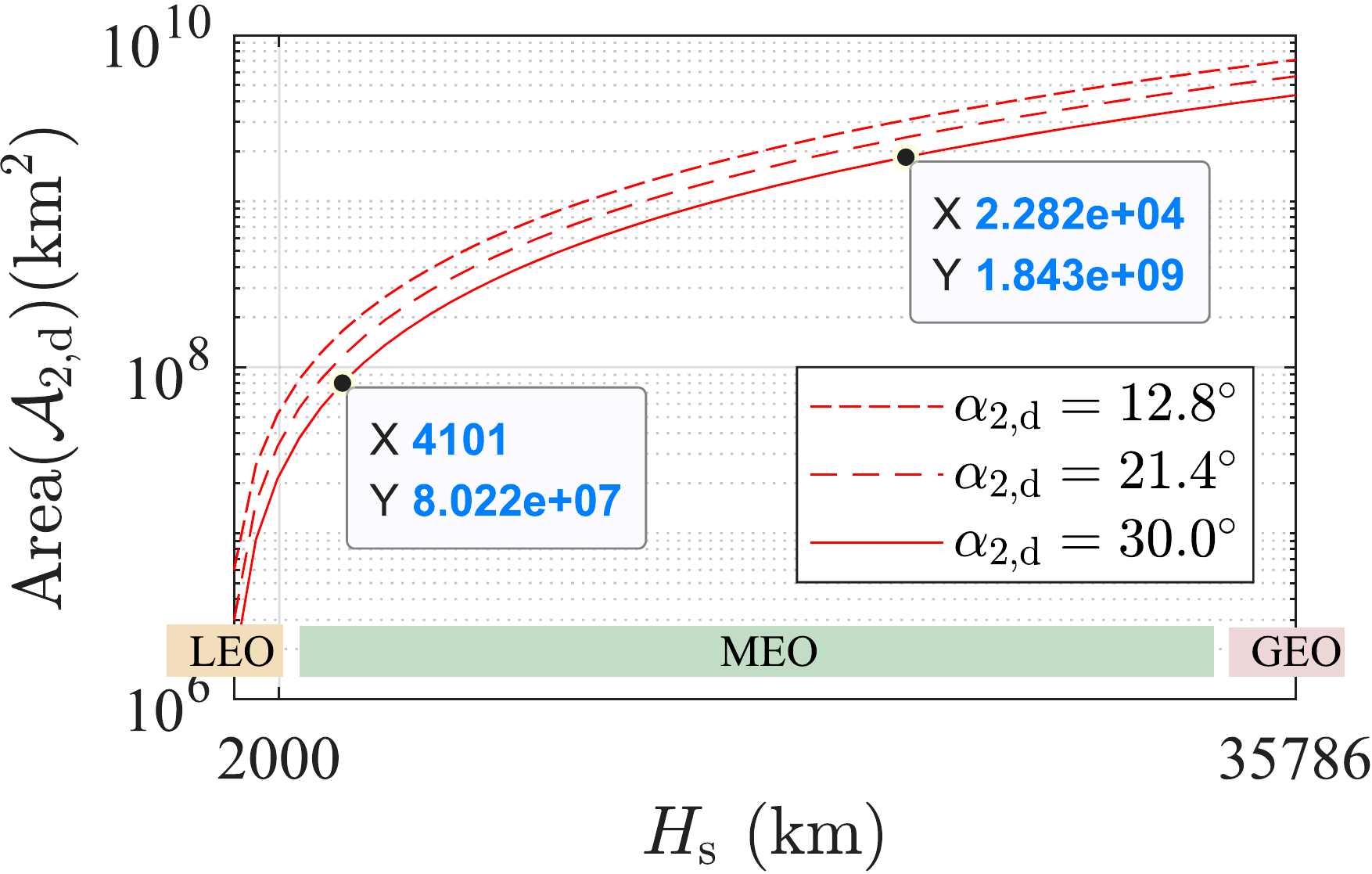}\label{fig: down2_area_H}}	\hfil	
	\\[-2ex]
	\subfloat[$\mathbb{S}_{3,\mathrm{u}}$(G2S)]{\includegraphics[width=4.5cm]{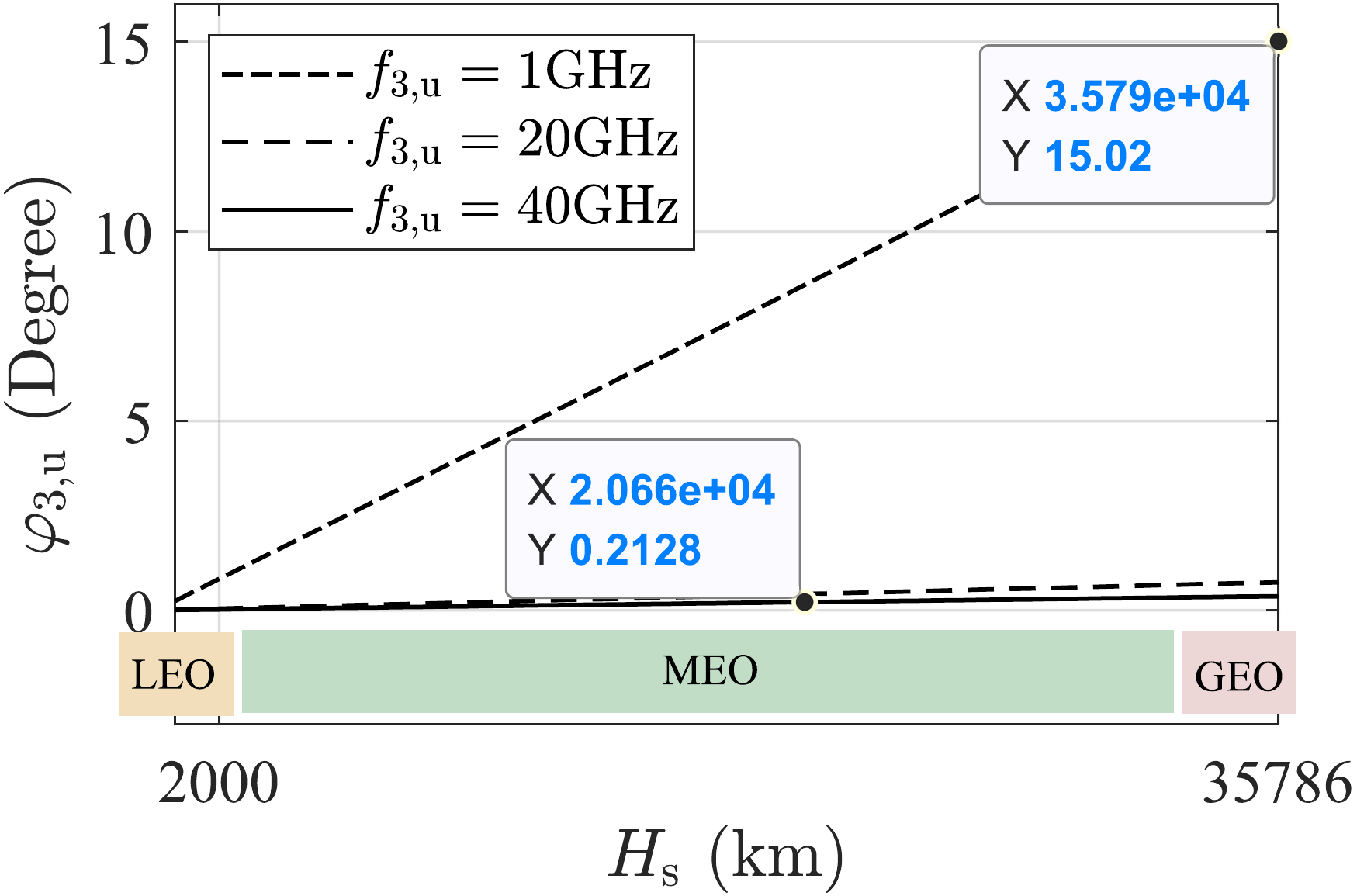}\label{fig: up3_phi_H}}	\hfil 
	\subfloat[$\mathbb{S}_{3,\mathrm{u}}$(G2S)]{\includegraphics[width=4.3cm]{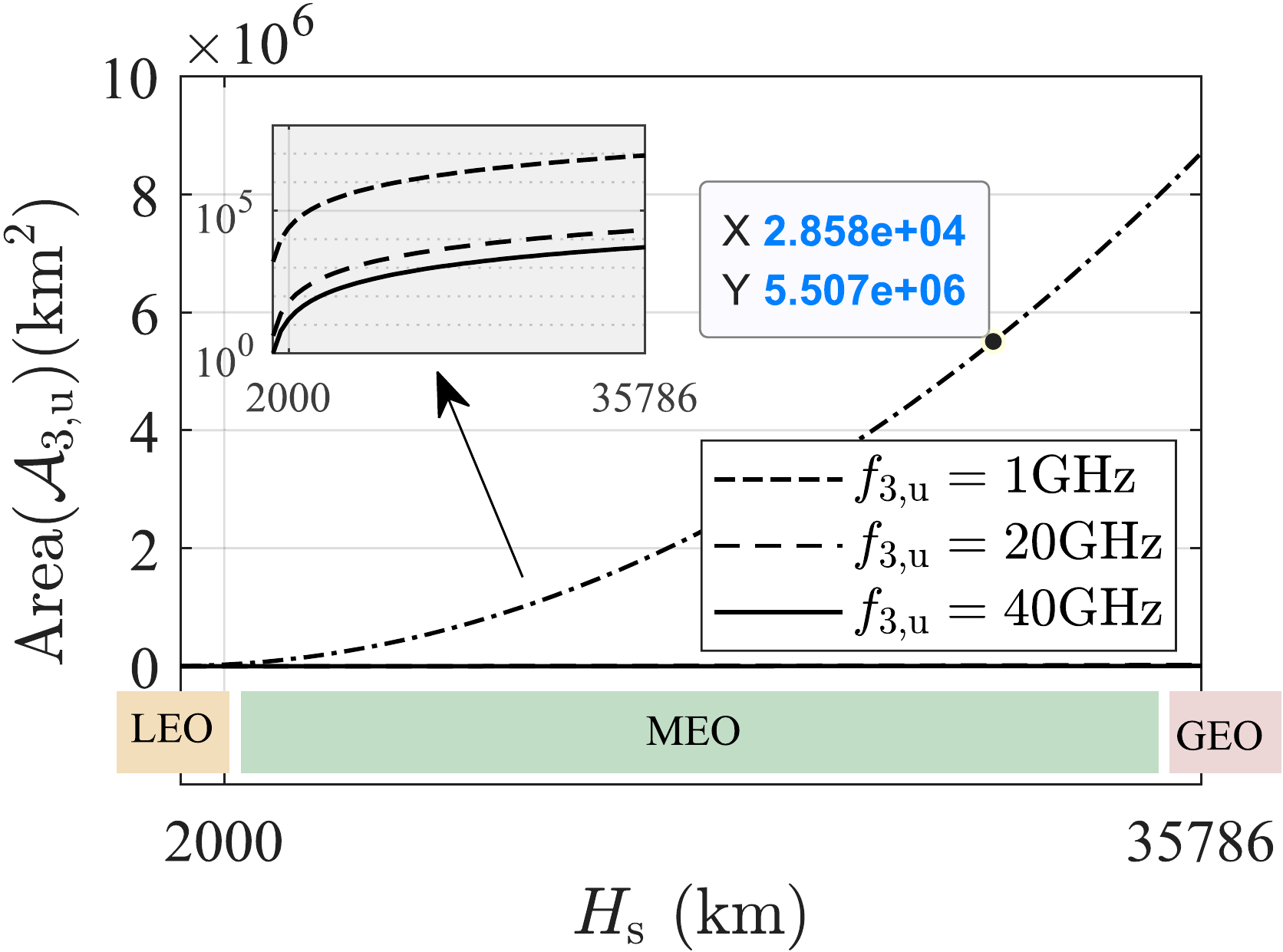}\label{fig: up3_area_H}}	\hfil
	\subfloat[$\mathbb{S}_{3,\mathrm{d}}$(S2G)]{\includegraphics[width=4.35cm]{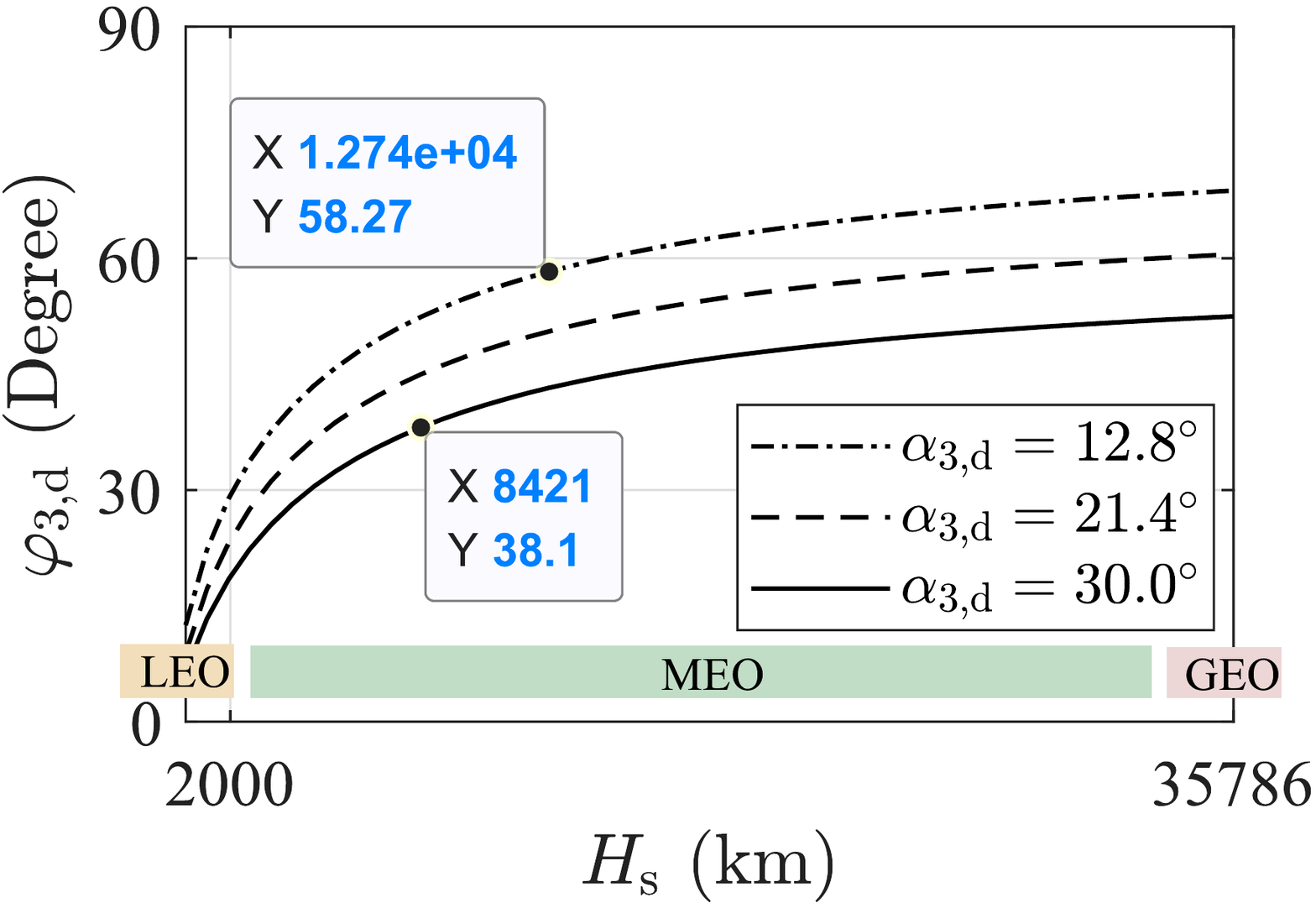}\label{fig: down3_phi_H}}	\hfil 
	\subfloat[$\mathbb{S}_{3,\mathrm{d}}$(S2G)]{\includegraphics[width=4.65cm]{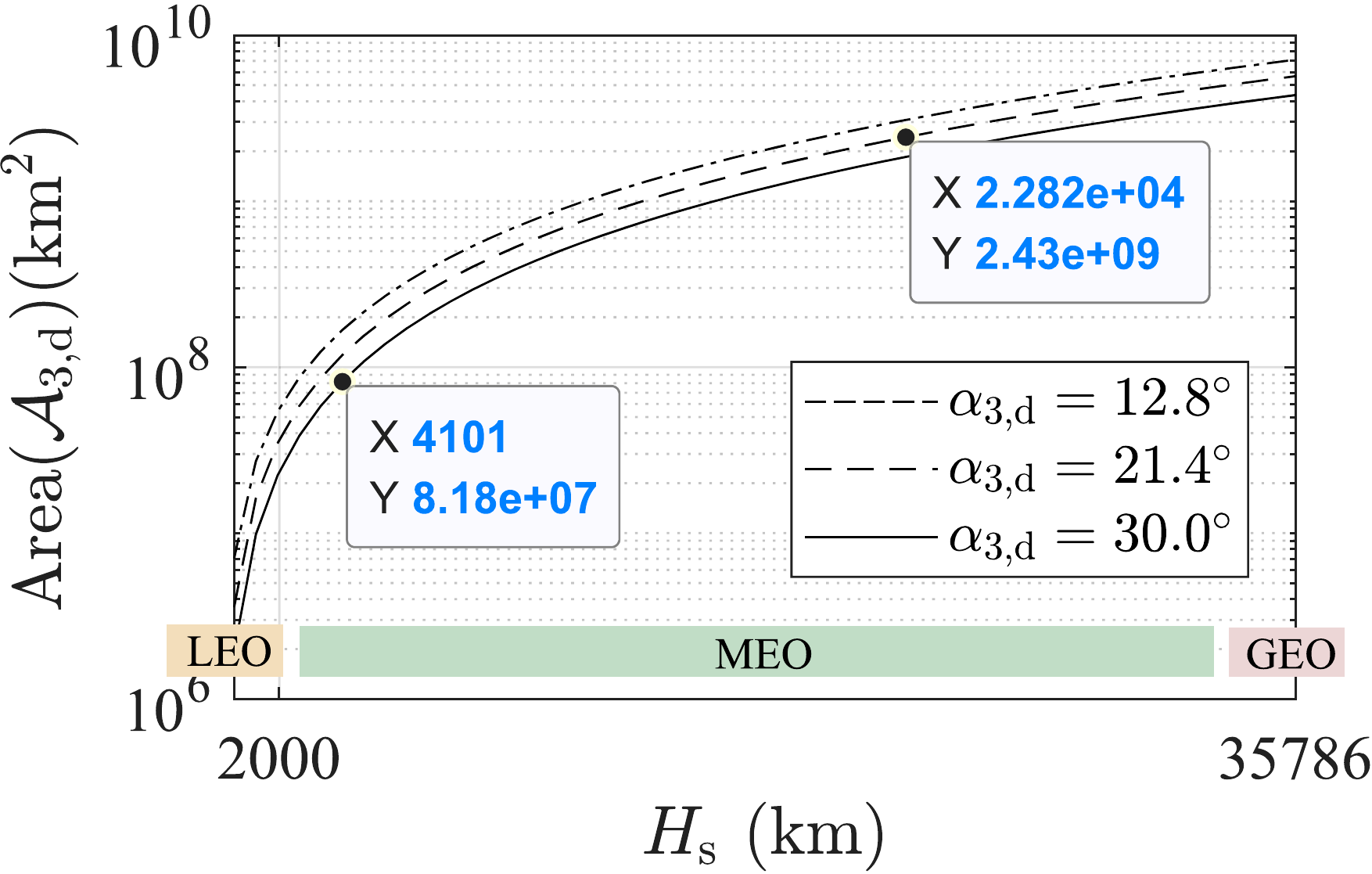}\label{fig: down3_area_H}}	\hfil
	\caption{The vertex angles $\varphi_{i,j}$ and the coverage area sizes $\mathrm{Area}(\mathcal{A}_{i,j})$ versus the carrier frequency $f_{i,\mathrm{u}}$, the receiver's elevation angle $\alpha_{i,\mathrm{d}}$, and the transmitter/receiver's height $H_t,H_r$ in six cross-layer scenarios $\mathbb{S}_{i,j}$ ($\forall i \in \{1,2,3\},j\in\{\mathrm{u},\mathrm{d}\}$). In $\mathbb{S}_{2,j}$ ($\forall j\in\{\mathrm{u},\mathrm{d}\}$), $H_{\mathrm{a}}=50$km. } 
	\label{fig: uplink}
	\vspace{-0.3cm}
\end{figure*}

As illustrated in Fig.~\ref{fig: uplink}, the coverage area $\mathrm{Area}(\mathcal{A}_{i,j})$ and the polar angle $\varphi_{i,j}$ in six cross-layer scenarios are effected by three key parameters: i) the carrier frequency, ii) the receiver's elevation angle, and iii) the transmitter/receiver's height. 

For three uplink scenarios $\mathbb{S}_{i,\mathrm{u}}$ ($\forall i \in \{1,2,3\}$), $\varphi_{i,\mathrm{u}}$ and $\mathrm{Area}(\mathcal{A}_{i,\mathrm{u}})$ decrease with the increasing of $f_{i,\mathrm{u}}$ and increase with the increasing of $H_r$. 
This is because the lower frequency enlarges the beamwidth, leading to the larger vertex angle, and the higher $H_r$ leads to wider coverage regions of the receiver. 
Among three uplink scenarios, 
$\varphi_{1,\mathrm{u}}$ 
is much smaller than $\varphi_{2,\mathrm{u}},\varphi_{3,\mathrm{u}}$. This is because of the much shorter altitude of aerial vehicles compared with the earth's radius. 
Moreover, comparing $\mathbb{S}_{2,\mathrm{u}}$ 
and $\mathbb{S}_{3,\mathrm{u}}$
, $\varphi_{3,\mathrm{u}}$ and $\mathrm{Area}(\mathcal{A}_{3,\mathrm{u}})$ are slightly larger than $\varphi_{2,\mathrm{u}}$ and $\mathrm{Area}(\mathcal{A}_{2,\mathrm{u}})$. Because $\mathbb{S}_{3,\mathrm{u}}$ has a longer 
transmission distance than $\mathbb{S}_{2,\mathrm{u}}$, bringing the larger region with a bigger vertex angle. 

For three downlink scenarios $\mathbb{S}_{i,\mathrm{d}}$ ($\forall i \in \{1,2,3\}$), $\varphi_{i,\mathrm{d}}$ and $\mathrm{Area}(\mathcal{A}_{i,\mathrm{d}})$ decrease with the increasing of $\alpha_{i,\mathrm{d}}$ and increase with the increasing of $H_r$. Because the larger elevation angle at the receiver leads to the narrower observation angle, resulting in the smaller coverage region. Meanwhile, the higher $H_r$ indicate wider observation regions from the receiver and larger vertex angles. 
In $\mathbb{S}_{1,\mathrm{d}}$ (i.e., the G2A scenario), $\varphi_{1,\mathrm{d}}$ is also very small compared with $\varphi_{2,\mathrm{d}}$ and $\varphi_{3,\mathrm{d}}$ because of the same reason in $\mathbb{S}_{1,\mathrm{u}}$. 
Furthermore, comparing two downlink scenarios $\mathbb{S}_{2,\mathrm{d}}$ 
and $\mathbb{S}_{3,\mathrm{d}}$, 
$\varphi_{3,\mathrm{d}}$ and $\mathrm{Area}(\mathcal{A}_{3,\mathrm{d}})$ are slightly larger than $\varphi_{2,\mathrm{d}}$ and $\mathrm{Area}(\mathcal{A}_{2,\mathrm{d}})$ because of the same reason in $\mathbb{S}_{2,\mathrm{d}}$ and $\mathbb{S}_{3,\mathrm{d}}$. 

\textbf{Insights:} The results above confirm our model's applicability across various parameters in all scenarios. Thus, our model can effectively characterize interfering node regions and transmission performance in SAGINs. Furthermore, the model can easily extend to practical networks by deploying different antennae and real-time coordinates of transceivers. E.g., the higher altitude of aerial vehicles in $\mathbb{S}_{2,j}$, the clearer difference among $\{\mathbb{S}_{2,j},\mathbb{S}_{3,j}\}$. 
\vspace{-0.8cm}

\section{Distributions under Spherical Coverage}
\label{sec: distri}
The unified coverage model can help to generate node distributions and facilitate network modeling in SAGINs. Accordingly, we design an algorithm to generate stochastic spatial coordinates of node distributions under spherical coverage areas. 
\vspace{-0.4cm}

\subsection{Algorithm to Generate Distributions}
\vspace{-0.1cm}
We present \textbf{\cref{algorithm1}} to generate node distributions for all six cross-layer scenarios. 
For $\mathbb{S}_{i,j}$, given a receiver $\mathrm{Rx}$, 
let $\Phi_t,\lambda_t$ be the stochastic distributions and the distribution density of transmitters covered by the receiver. Taking the Poisson point process (PPP) distribution as an example, \textbf{\cref{algorithm1}} generate $\Phi_t\in\mathcal{A}_{i,j}$ under the spherical coverage region $\mathcal{A}_{i,j}$. Changing PPP in \cref{line5,line6,line7} of \textbf{\cref{algorithm1}}, our algorithm can be extended to other distribution models for practical use.

\begin{figure*}[t]
	\centering 
	\subfloat[$\mathbb{S}_{1,\mathrm{u}}\mathrm{(G2A)},\mathbb{S}_{2,\mathrm{u}}\mathrm{(A2S)},\mathbb{S}_{3,\mathrm{u}}\mathrm{(G2S)}$]{\includegraphics[width=8.7cm]{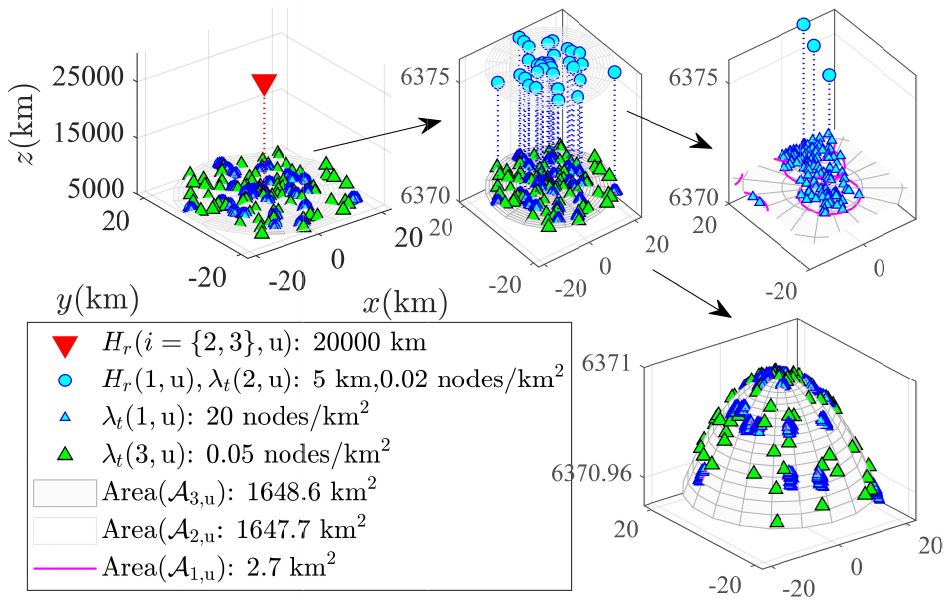}\label{fig: up_simu0}} 	\hfil 
	\subfloat[$\mathbb{S}_{1,\mathrm{d}}\mathrm{(A2G)},\mathbb{S}_{2,\mathrm{d}}\mathrm{(S2A)},\mathbb{S}_{3,\mathrm{d}}\mathrm{(S2G)}$]{\includegraphics[width=8.7cm]{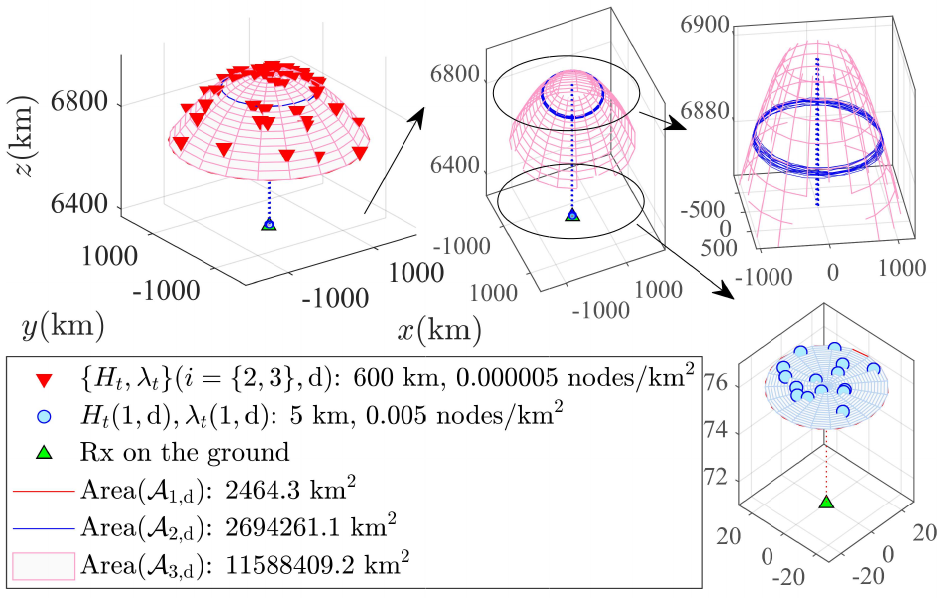}\label{fig: down_simu0}} 	\hfil 
	\caption{Simulations of node distributions in six cross-layer scenarios. 
		In (a), $f_{1,\mathrm{u}}=2\mathrm{GHz},f_{2,\mathrm{u}}=40\mathrm{GHz},f_{3,\mathrm{u}}=40\mathrm{GHz}$. 
		In (b), 
		$\alpha_{2,\mathrm{d}}=10^{\circ},\alpha_{2,\mathrm{d}}=30^{\circ},\alpha_{2,\mathrm{d}}=10^{\circ}$. In (b), the enlarged result of $\mathbb{S}_{2,\mathrm{d}}$ hides the satellite distribution to show $\mathcal{A}_{2,\mathrm{d}}$ from multiple aerial vehicles.} 
	\label{fig: simu}
	\vspace{-0.3cm}
\end{figure*}
In \textbf{\cref{algorithm1}}, 
the input is initial system parameters 
in $\mathbb{S}_{i,j}$, including the earth radius $R_e$, the light speed $c$, the illumination coefficient $\kappa_{i,\mathrm{u}}$ and the diameter $\mathrm{D}_{i,\mathrm{u}}$ of the receiver antenna (for uplink scenarios), the carrier frequency $f_{i,\mathrm{u}}$ (for uplink scenarios), the minimum elevation angle $\alpha_{i,\mathrm{d}}$ of the receiver (for downlink scenarios), the transmitter and receiver altitudes $H_t,H_r$, the distribution density of transmitters $\lambda_t$, and the azimuth and polar angles $\vartheta_r,\varphi_r$ of the receiver.~\cref{uplink01,uplink1,uplink3,uplink02,downlink1,uplink03,arealine} 
calculates $R_t,R_r,\delta_{i,j}$ and $\mathrm{Area}(\mathcal{A}_{i,\mathrm{u}})$. 
\cref{line5,line6,line7} generate a PPP distribution $\Phi_t=\{(\vartheta_{t,\mathrm{ppp}},\varphi_{t,\mathrm{ppp}})\}$ centered at $(0,0)$, where $\mathbf{n}_{\mathrm{ppp}}$ is the random PPP parameter that means the transmitters' amount on $\mathcal{A}_{i,\mathrm{u}}$; \cref{line8} constructs a yaw-pitch matrix $\mathbf{yp}_r$ to rotate the center of $\Phi_t$ to $(\vartheta_r,\varphi_r)$. $\mathbf{yp}_r$ is 
given by
\vspace{-0.4cm}

{\footnotesize
	\begin{align}
		&\mathbf{yp}_r=\begin{bmatrix}
			\cos(\vartheta_r)&-\sin(\vartheta_r)  & 0\\ 
			\sin(\vartheta_r)&\cos(\vartheta_r)  & 0 \\ 
			0&  0& 1
		\end{bmatrix}\times\begin{bmatrix}
			\cos(\varphi_r)&0 & \sin(\varphi_r) \\ 
			0&1 & 0 \\ 
			-\sin(\varphi_r)&  0& \cos(\varphi_r)
		\end{bmatrix}.\label{eq: yp_r}
	\end{align}
}%
\vspace{-0.4cm}

\noindent From \cref{line9} to \cref{line19}, the algorithm generates the $x,y,z$ coordinates of the transmitter distribution centered at $(\vartheta_r,\varphi_r)$ and distributed on the spherical surface with the radius $R_t$. 
The algorithm complexity arises from the calculation of~\cref{cor1} and node generations. The former expressions can be easily computed by substituting system parameters, and the latter needs more computing power as the increasing nodes.

\begin{adjustwidth}{0cm}{0.1cm}
	\begin{algorithm}[tb]
		\SetKwInOut{Input}{Input}
		\SetKwInOut{Output}{Output}
		\Input{$R_e,c,\kappa_{i,\mathrm{u}},\mathrm{D}_{i,\mathrm{u}},f_{i,\mathrm{u}},\alpha_{i,\mathrm{d}},H_t,H_r,\lambda_t,\vartheta_r,\varphi_r$}
		\Output{$\Phi_t\in\mathcal{A}_{i,j}$}
		$R_t\gets R_e+H_t$; $R_r\gets R_e+H_r$ \label{line0} \\
		\eIf{$j=\mathrm{u}$}{\label{uplink01}
			$\theta_{i,\mathrm{u}}\gets$substituting $c,\kappa_{i,\mathrm{u}},\mathrm{D}_{i,\mathrm{u}},f_{i,\mathrm{u}}$ into \eqref{eq: theataG} \label{uplink1}\\
			$\delta_{i,j}\gets\delta_{i,\mathrm{u}}\gets$substituting $\theta_{i,\mathrm{u}},R_t,R_r$ into~\cref{cor1} \label{uplink3}\\
		}
		{\label{uplink02}
			$\delta_{i,j}\gets\delta_{i,\mathrm{d}}\gets$substituting $\alpha_{i,\mathrm{d}},R_t,R_r$ into~\cref{cor1} \label{downlink1}\\
		}
		{\label{uplink03}		
			$\mathrm{Area}(\mathcal{A}_{i,j})\gets$substituting $R_t,\varphi_{i,j}$ into~\cref{the1} \label{arealine} \\		
			$\mathbf{n}_{\mathrm{ppp}}\gets\mathrm{poissrnd}\left(\left \lfloor \lambda_t*\mathrm{Area}(\mathcal{A}_{i,j}) \right \rfloor\right)$ \label{line5}\\
			$\vartheta_t\gets 2*\pi *\mathrm{rand}(\mathbf{n}_{\mathrm{ppp}},1)$\label{line6}\\
			$\varphi_t\gets \varphi_{i,j} *(2*\mathrm{rand}(\mathbf{n}_{\mathrm{ppp}},1)-1)$ \label{line7}\\
			$\mathbf{yp}_r\gets$substituting $\vartheta_r,\varphi_r$ into \eqref{eq: yp_r} \label{line8}\\		
			$x_t\gets []$, $y_t\gets []$,
			$z_t\gets []$ \label{line9}\\
		}
		\For{$i\gets 1$ \KwTo $\mathbf{n}_{\mathrm{ppp}}$}{ 
			$z_0\gets R_t*\cos(\varphi_t(i))$ \label{line11} \\
			$x_0\gets R_t*\sin(\varphi_t(i))*\cos(\vartheta_t(i))$ \label{line12} \\
			$y_0\gets R_t*\sin(\varphi_t(i))*\sin(\vartheta_t(i))$ \label{line13}\\
			$\mathbf{rotation}\gets\mathbf{yp}_r*[z_0;x_0;y_0]$ \label{line14}\\
			$x_t\gets[x_t,\mathbf{rotation}(1)]$ \label{line15}\\
			$y_t\gets[y_t,\mathbf{rotation}(2)]$ \label{line16}\\
			$z_t\gets[z_t,\mathbf{rotation}(3)]$ \label{line17}\\
		}
		$\Phi_t\gets \{(x_t,y_t,z_t)\}$ \label{line19} 
		\caption{The algorithm to generate $\Phi_t\in\mathcal{A}_{i,j}$}
		\label{algorithm1}
	\end{algorithm}
\end{adjustwidth}
\vspace{-0.3cm}

\subsection{Simulation Results}
\vspace{-0.1cm}
Fig.~\ref{fig: simu} shows MATLAB simulation results of the transmitters' distribution of SAGINs 
following \textbf{\cref{algorithm1}}. 

Fig.~\ref{fig: simu}\subref{fig: up_simu0} shows three uplink scenarios under a reference satellite with $H_{\mathrm{s}}=20000\mathrm{km}$ (i.e., the MEO orbit) and aerial vehicles flying at $H_{\mathrm{a}}=5\mathrm{km}$. The coverage region areas from the satellite in $\mathbb{S}_{2,\mathrm{u}}$ and $\mathbb{S}_{3,\mathrm{u}}$ are $1648.6\mathrm{km}^2$ on ground and $1647.7\mathrm{km}^2$ on air, respectively. Giving a sparse density of ground and air transmitters ($0.05\mathrm{nodes}/\mathrm{km}^2$ and $0.02\mathrm{nodes}/\mathrm{km}^2$), the satellite still covers a lot transmitters (i.e., $0.05\times1648.6=82.43\mathrm{nodes}$ and $0.02\times1647.7=32.954\mathrm{nodes}$). Comparatively, the ground coverage region from an aerial vehicle (in $\mathbb{S}_{1,\mathrm{u}}$) is much smaller but it can support the equal and even more transmitters (i.e., $20\times2.7=54\mathrm{nodes}$) efficiently. Considering the uplink transmissions via ground-to-air-to-satellite path, a rough calculation of supported transmitters are $54\times32.954=1,779.516\mathrm{nodes}$, which are more than a single group-to-satellite path ($82.43\mathrm{nodes}$). 
\vspace{-0.1cm}

Fig.~\ref{fig: simu}\subref{fig: down_simu0} shows three downlink scenarios under a reference ground user where satellites are located at $H_{\mathrm{s}}=600\mathrm{km}$ (i.e., the LEO orbit) and aerial vehicles flying at $H_{\mathrm{a}}=5\mathrm{km}$. We simulate a PPP distribution of satellites with the density $0.000005\mathrm{nodes}/\mathrm{km}^2$, corresponding to $4\pi (R_e+H_{\mathrm{s}})^2\times 0.000005\mathrm{nodes}/\mathrm{km}^2\approx 3,053$ satellites deployed in the LEO orbit. The coverage region areas on satellites in $\mathbb{S}_{2,\mathrm{d}}$ and $\mathbb{S}_{2,\mathrm{d}}$ are $11588409.2\mathrm{km}^2$ from a ground node and $2694261.1\mathrm{km}^2$ from an aerial vehicle, respectively. Then the observed number of satellites in $\mathbb{S}_{2,\mathrm{d}}$ and $\mathbb{S}_{3,\mathrm{d}}$ can be roughly calculated, i.e., $11588409.2\times 0.000005\approx 58\mathrm{nodes}$ from a ground node and $2694261.1\times 0.000005\approx 13\mathrm{nodes}$, which can be further used to analyze the downlink capacity in SAGINs. In $\mathbb{S}_{1,\mathrm{d}}$, the coverage region from ground to air is much smaller (i.e., $2464.3\mathrm{km}^2$) and it can observe $0.005\times2464.3\approx12$ aerial vehicles with the given aerial-vehicle density $0.005 \mathrm{nodes}/\mathrm{km}^2$. 
It can be seen that the coverage regions on the space layer from several aerial vehicles 
are nearly coincided. This is because i) all aerial vehicles on $\mathcal{A}_{1,\mathrm{d}}$ are nearly a dot compared with the long propagation distance to the satellite, and ii) the large elevation angle chosen in aerial vehicles $\alpha_{2,\mathrm{d}}=30^{\circ}$. The aerial vehicle relayed downlink path won't have the enhanced performance as in uplink scenarios.
\vspace{-0.1cm}

\textbf{Insights:} 
Our models and observations offer the following implementation insights: 1) The unified vertex angle within the coverage area can be used to model node distributions in SAGINs and facilitate large-scale network modeling. 2) The coverage area of the six cross-layer scenarios in SAGINs can be enhanced by increasing the receiver height (for all six scenarios), decreasing the carrier frequency (for three uplink scenarios), or reducing the receiver's elevation angle (for three downlink scenarios). 3) According to our numerical results, downlink scenarios feature higher altitudes and wider observation angles at the receiver (due to auto-tracking systems) compared to uplink scenarios, resulting in a larger coverage area for downlink scenarios. Understanding this coverage area difference can aid in the deployment of network nodes, such as determining their density and distribution, thereby coordinating downlink and uplink capacity.
\vspace{-0.4cm}

\section{Conclusion}
\label{sec: con}
\vspace{-0.1cm}
In this paper, we develop a unified analytical model of coverage regions for six cross-layer scenarios in SAGINs. Using spherical geometry for both uplink and downlink scenarios, we derive analytical formulas for the vertex angles and area sizes of coverage regions across all six cross-layer scenarios. 
We conducted extensive numerical results to examine the coverage regions under varying carrier frequencies and receivers' altitudes for six scenarios. Moreover, we designed an algorithm to generate stochastic distributions of transmitters within spherical coverage regions. Finally, we simulated and visualized the accurate spatial coordinates of all distributed transmitters. Our presented model and algorithm have great potential to assist in network modeling and testing of SAGINs before practical deployments. 
\vspace{-0.4cm}

\section*{Acknowledgment}
\vspace{-0.1cm}
\begin{spacing}{0.7}
	{
		\footnotesize
		The work in this paper was supported in part by the Hong Kong Metropolitan University Research Grant No. RD/2023/2.22, in part by two grants from the Research Grants Council of the Hong Kong Special Administrative Region, China, under projects No. UGC/FDS16/E15/24 and No. UGC/FDS16/E02/22, in part by the Hong Kong Research Matching Grant No. CP/2022/2.1 in the Central Pot, and in part by the Team-based Research Fund No. TBRF/2024/1.10.
	}
\end{spacing}
\vspace{-0.4cm}

\section*{Appendix~A}
\label{app: angleproof1}
\vspace{-0.1cm}
\small
\textit{The proof of Corollary}~\ref{cor1}: 
Follow the spherical geometry in Fig.~\ref{fig: system}\subref{subfig: uplink}, the calculation of $\varphi_{i,\mathrm{u}}$ includes two cases below. 

\textit{Case 1: } When $\theta_{i,\mathrm{u}}/2> \arcsin(R_t/R_r)$, 
drawing a group of tangent lines from $\mathrm{Rx}$ on $\mathcal{O}(\mathrm{Tx})$, all tangent points form the edge of $\mathcal{A}_{i,\mathrm{u}}$. We have $\varphi_{i,\mathrm{u}}=\measuredangle\mathrm{RxOTx_e}=\arccos\left(R_t/R_r\right)$. 
%


\textit{Case 2: } When $\theta_{i,\mathrm{u}}/2\leq \arcsin(R_t/R_r)$, 
drawing a group of lines from the beam surface of $\mathrm{Rx}$ on $\mathcal{O}(\mathrm{Tx})$, all these lines from $\mathrm{Rx}$ on $\mathcal{O}(\mathrm{Tx})$ form the edge of $\mathcal{A}_{i,\mathrm{u}}$, we have 
\vspace{-0.5cm}

{\footnotesize
	\begin{align}
		&\cos\left (\measuredangle\mathrm{RxOTx_e} \right )=\cos \varphi_{i,\mathrm{u}} =\frac{R_t^2+R_r^2-\mathrm{RxTx_e}^2}{2R_tR_r },\label{eq: cosSOP}\\
%
		&\cos\left (\measuredangle\mathrm{ORxTx_e} \right )=\cos\left ( \frac{\theta_{i,\mathrm{u}}}{2} \right )
		=\frac{\mathrm{RxTx_e}^2+R_r^2-R_t^2}{2(\mathrm{RxTx_e})R_r }.\label{eq: cosOSP}
	\end{align}%
}%
\vspace{-0.3cm}

\noindent Then $\mathrm{RxTx_e}=\cos\left ( \theta_{i,\mathrm{u}}/2 \right )R_r \pm \sqrt{R_t^2-\sin^2\left ( \theta_{i,\mathrm{u}}/2 \right )R_r^2}$.
Since $\measuredangle\mathrm{RxOTx_e}$ is an acute angle in $\bigtriangleup \mathrm{RxOTx_e}$, $\pm$ is reduced to $-$. 
Substituting $\mathrm{RxTx_e}$ into~\cref{eq: cosSOP}, we have 
\vspace{-0.5cm}

{\footnotesize
	\begin{align}
		&\varphi_{i,\mathrm{u}} = \frac{R_r}{R_t}\sin^2\left(\frac{\theta_{i,\mathrm{u}}}{2}\right) 
		+\cos\left (\frac{\theta_{i,\mathrm{u}}}{2} \right )\sqrt{1-\left (\frac{R_r}{R_t}\sin\left(\frac{\theta_{i,\mathrm{u}}}{2}\right )\right )^2}.\label{eq: varSoA2}
	\end{align}
}%
\vspace{-0.3cm}

\noindent To sum up, we have the results of $\varphi_{i,\mathrm{u}}$ in Corollary~\ref{cor1}. Follow the spherical geometry in Fig.~\ref{fig: system}\subref{subfig: downlink}, $\varphi_{i,\mathrm{d}}$ can be calculated by
\vspace{-0.5cm}

{\footnotesize
	\begin{align}
		&\cos \varphi_{i,\mathrm{d}}=\cos\left(\measuredangle\mathrm{RxOTx_e}\right)=\frac{R_r^2+R_t^2-\mathrm{RxTx_e}^2}{2R_rR_t }, \label{eq: cosRxOTx}\\
%
		&\cos\left(\measuredangle\mathrm{Tx}_{\perp}\mathrm{Tx_eRx}\right)=\cos \alpha_{i,\mathrm{d}}=\frac{\mathrm{Tx}_{\perp}\mathrm{Tx_e}}{\mathrm{RxTx_e}}=\frac{\sin\varphi_{i,\mathrm{d}}\times \mathrm{OTx_e}}{\mathrm{RxTx_e}}, \label{eq: P2prepA}
	\end{align}%
}%
\vspace{-0.3cm}

\noindent where \eqref{eq: cosRxOTx} is resulted from Laws of Cosines in $\bigtriangleup\mathrm{RxOTx_e}$. 
Drawing an auxiliary vertical line $\mathrm{Tx_eTx}_{\perp}$ from the point $\mathrm{Tx_e}$ to the line $\mathrm{ORx}$ and letting $\mathrm{Tx}_{\perp}$ be the vertical point at the line $\mathrm{ORx}$, then \eqref{eq: P2prepA} is resulted from two right triangles $\bigtriangleup\mathrm{RxTx_{\perp} Tx_e}$ and $\bigtriangleup\mathrm{OTx_{\perp} Tx_e}$. Substituting $\mathrm{OTx_e}=R_t,\mathrm{ORx}=R_r$ into \eqref{eq: P2prepA}, we have $\mathrm{RxTx_e}$. Combining the results of $\mathrm{RxTx_e}$ in \eqref{eq: cosRxOTx} and \eqref{eq: P2prepA}, we have 
\vspace{-0.4cm}

{\footnotesize
	\begin{align}
		\cos\varphi_{i,\mathrm{d}} = \frac{R_r}{R_t}\cos^2\alpha_{i,\mathrm{d}} \pm \sin\alpha_{i,\mathrm{d}} \sqrt{1-\left(\frac{R_r}{R_t}\cos^2\alpha_{i,\mathrm{d}}\right)^2}.\label{eq: phi2''}
	\end{align}
}%
\vspace{-0.3cm}

\noindent 
Since $\measuredangle\mathrm{RxOTx_e}$ is an acute angle in $\bigtriangleup \mathrm{RxOTx_e}$, $\pm$ in~\cref{eq: phi2''} is reduced to $+$. To sum up, we have 
Corollary~\ref{cor1}. 
\hfill$\blacksquare$	
\vspace{-0.4cm}

\balance
\bibliographystyle{IEEEtran}
\bibliography{ref}	

\begin{thebibliography}{10}
\providecommand{\url}[1]{#1}
\csname url@samestyle\endcsname
\providecommand{\newblock}{\relax}
\providecommand{\bibinfo}[2]{#2}
\providecommand{\BIBentrySTDinterwordspacing}{\spaceskip=0pt\relax}
\providecommand{\BIBentryALTinterwordstretchfactor}{4}
\providecommand{\BIBentryALTinterwordspacing}{\spaceskip=\fontdimen2\font plus
\BIBentryALTinterwordstretchfactor\fontdimen3\font minus \fontdimen4\font\relax}
\providecommand{\BIBforeignlanguage}[2]{{%
\expandafter\ifx\csname l@#1\endcsname\relax
\typeout{** WARNING: IEEEtran.bst: No hyphenation pattern has been}%
\typeout{** loaded for the language `#1'. Using the pattern for}%
\typeout{** the default language instead.}%
\else
\language=\csname l@#1\endcsname
\fi
#2}}
\providecommand{\BIBdecl}{\relax}
\BIBdecl

\bibitem{3GPPTR38863}
\BIBentryALTinterwordspacing
3GPP, ``{Solutions for NR to support non-terrestrial networks (NTN) (Release 17)},'' 2023-03, accessed on April 9, 2024. [Online]. Available: \url{https://www.3gpp.org/ftp/Specs/archive/38_series/38.863/}
\BIBentrySTDinterwordspacing

\bibitem{23GSA}
\BIBentryALTinterwordspacing
GSA, ``{5G Satellite-Connectivity August 2023},'' Aug., 2023, accessed on Feb. 10, 2024. [Online]. Available: \url{https://gsacom.com/paper/5g-satellite-connectivity-august-2023/}
\BIBentrySTDinterwordspacing

\bibitem{3GPPTR38811}
\BIBentryALTinterwordspacing
3GPP, ``{Study on New Radio (NR) to support non-terrestrial networks (Release 15)},'' 2020-09, accessed on April 9, 2024. [Online]. Available: \url{https://www.3gpp.org/ftp/Specs/archive/38_series/38.811/}
\BIBentrySTDinterwordspacing

\bibitem{24iotleo}
A.~K. Dwivedi, S.~Chaudhari, N.~Varshney, and P.~K. Varshney, ``{Performance Analysis of LEO Satellite-Based IoT Networks in the Presence of Interference},'' \emph{IEEE Internet of Things Journal}, vol.~11, no.~5, pp. 8783--8799, 2024.

\bibitem{23BPPsagin}
Q.~Chen, W.~Meng, S.~Han, C.~Li, and T.~Q.~S. Quek, ``{Coverage Analysis of SAGIN With Sectorized Beam Pattern Under Shadowed-Rician Fading Channels},'' \emph{IEEE Transactions on Communications}, vol.~71, no.~8, pp. 4988--5004, 2023.

\bibitem{21twcsg}
Y.~Tian, G.~Pan, M.~A. Kishk, and M.-S. Alouini, ``{Stochastic Analysis of Cooperative Satellite-UAV Communications},'' \emph{IEEE Transactions on Wireless Communications}, vol.~21, no.~6, pp. 3570--3586, 2022.

\bibitem{zhang2021stochastic}
X.~Zhang, B.~Zhang, K.~An, G.~Zheng, S.~Chatzinotas, and D.~Guo, ``{Stochastic Geometry-Based Analysis of Cache-Enabled Hybrid Satellite-Aerial-Terrestrial Networks With Non-Orthogonal Multiple Access},'' \emph{IEEE Transactions on Wireless Communications}, vol.~21, no.~2, pp. 1272--1287, 2022.

\bibitem{al2022optimal}
B.~Al~Homssi and A.~Al-Hourani, ``{Optimal Beamwidth and Altitude for Maximal Uplink Coverage in Satellite Networks},'' \emph{IEEE Wireless Communications Letters}, 2022.

\bibitem{liu2023TVT}
Y.~Liu, Q.~Wang, H.-N. Dai, Y.~Fu, N.~Zhang, and C.~C. Lee, ``{UAV-Assisted Wireless Backhaul Networks: Connectivity Analysis of Uplink Transmissions},'' \emph{IEEE Transactions on Vehicular Technology}, pp. 1--13, 2023.

\bibitem{talgat2024stochastic}
A.~Talgat, M.~A. Kishk, and M.-S. Alouini, ``{Stochastic geometry-based uplink performance analysis of IoT over LEO satellite communication},'' \emph{IEEE Transactions on Aerospace and Electronic Systems}, 2024.

\bibitem{liu2024}
Y.~Liu, H.-N. Dai, Q.~Wang, O.~J. Pandey, Y.~Fu, N.~Zhang, D.~Niyato, and C.~C. Lee, ``{Space-Air-Ground Integrated Networks: Spherical Stochastic Geometry-Based Uplink Connectivity Analysis},'' \emph{IEEE Journal on Selected Areas in Communications}, vol.~42, no.~5, pp. 1387--1402, 2024.

\bibitem{maral2020satellite}
G.~Maral, M.~Bousquet, and Z.~Sun, \emph{{Satellite communications systems: systems, techniques and technology}}.\hskip 1em plus 0.5em minus 0.4em\relax John Wiley \& Sons, 2020.

\bibitem{173GPP}
\BIBentryALTinterwordspacing
3GPP, ``{Study on Enhanced LTE Support for Aerial Vehicles (Release 15)},'' 2017-12, accessed on April 9, 2024. [Online]. Available: \url{https://www.3gpp.org/ftp/Specs/archive/36_series/36.777/}
\BIBentrySTDinterwordspacing

\bibitem{3GPPTR381015}
\BIBentryALTinterwordspacing
------, ``{User Equipment (UE) radio transmission and reception - Part 5: Satellite access Radio Frequency (RF) and performance requirements},'' 2024-03, accessed on April 9, 2024. [Online]. Available: \url{https://www.3gpp.org/ftp/Specs/archive/38_series/38.101/}
\BIBentrySTDinterwordspacing

\end{thebibliography}
\end{document}